\documentclass[english,global]{svjour}
\usepackage[T1]{fontenc}
\usepackage[latin9]{inputenc}
\usepackage{geometry}
\usepackage{float}
\usepackage{url}
\usepackage{amsmath}
\usepackage{graphicx}
\usepackage{amssymb}

\providecommand{\tabularnewline}{\\}

\journalname{}
\usepackage{epsf}
\usepackage{rotate}
\usepackage{epsfig} 
\usepackage{graphicx} 
\usepackage{graphics} 

\usepackage{babel}

\begin{document}

\title{Statistical Mechanics of Two Hard Spheres in a Spherical Pore,\\
Exact Analytic Results in $D$ Dimension}

\author{Ignacio Urrutia and Leszek Szybisz}

\institute{Departamento de F\'{i}sica, Comisi\'on Nacional de Energ\'{i}a
At\'omica, Av. Gral. Paz 1499 (RA-1650) San Mart\'{i}n, Buenos Aires,
Argentina. Departamento de F\'{i}sica, Facultad de Ciencias Exactas
y Naturales, Universidad de Buenos Aires, Ciudad Universitaria, RA-1428
Buenos Aires, Argentina. Member of the carrera del investigador, CONICET. }

\mail{iurrutia@cnea.gov.ar}

\maketitle

\abstract{This work is devoted to the exact statistical mechanics treatment
of simple inhomogeneous few-body systems. The system of two Hard Spheres
(HS) confined in a hard spherical pore is systematically analyzed
in terms of its dimensionality $D$. The canonical partition function,
and the one- and two-body distribution functions are analytically
evaluated and a scheme of iterative construction of the $D+1$ system
properties is presented. We analyse in detail both the effect of high
confinement, when particles become caged, and the low density limit.
Other confinement situations are also studied analytically and several
relations between, the two HS in a spherical pore, two sticked HS
in a spherical pore, and two HS on a spherical surface partition functions
are traced. These relations make meaningful the limiting caging and
low density behavior. Turning to the system of two HS in a spherical
pore, we also analytically evaluate the pressure tensor. The thermodynamic
properties of the system are discussed. To accomplish this statement
we purposely focus in the overall characteristics of the inhomogeneous
fluid system, instead of concentrate in the peculiarities of a few
body system. Hence, we analyse the equation of state, the pressure
at the wall, and the fluid-substrate surface tension. The consequences
of new results about the spherically confined system of two HS in
D dimension on the confined many HS system are investigated. New constant
coefficients involved in the low density limit properties of the open
and closed system of many HS in a spherical pore are obtained for
arbitrary $D$. The complementary system of many HS which surrounds
a hard sphere (a cavity inside of a bulk HS system) is also discussed.}

\section{Introduction\label{sec:Introduction}}

The Hard Spheres (HS) and Hard Disks (HD) systems have attracted the
interest of many people, because they constitute prototypical simple
fluids and even solids \cite{Alder,Barker,eos_HD,Lowen,Sals,eos_HS,Wood}.
The extension to arbitrary dimensionality of this hard spherical particle
system has been the object of several studies too \cite{Baus,B4_evenD,B8_D,Loes,Luban,B4_oddD,Wyler}.
Though the apparent simplicity of these systems, only a few exact
analytical results for the homogeneous system are known concerning
mainly the dimensional dependence of the first four virial coefficients
\cite{B4_evenD,Luban1,B4_oddD} existing numerical extensions to higher
order \cite{B8_D,Krat}. Exact virial series studies has been also
done in inhomogeneous systems \cite{Bell,McQ,Soko} in three dimensions.
Recently some attention was dedicated to very small and inhomogeneous
systems of HS and HD confined in small vessels. The study of small
systems constrained to differently shaped cavities has enlightening
aspects of loss of ergodicity, glass transitions, thermodynamic second
law, and some other fundamental questions of statistical mechanics
and thermodynamic \cite{HD_2inBoxMD,2ndlaw,HSp_2inSphMC,Nemeth,HD_2inBoxMDbis,HDp_2inBoxMD,Erg}.

The exact evaluation of the properties for continuous (non-lattice)
inhomogeneous systems of few particles is a new trend in statistical
mechanics. The analyzed systems are usually HS and HD where the hard
potential represents the simplest non null interaction and different
ensembles approach may be done. An indirect result of these calculations
is the exact volume, size and even number of particle dependence of
low order cluster integrals \cite{Bell,BookHill,Kratky,McQ,Urru}.
Until now, the systems of two and three equal HDs in a rectangular
box has been solved \cite{HD_3inBox,HD_2inBox} more recently two
HS in a box \cite{HS_2inBox} and two HS and HD in a hollow or spherical
cavity \cite{Urru} were studied. A systematic dimensional approach
of the two HS in a box system was also done in \cite{HS_2inBox},
where an iterative construction framework was adopted for the increasing
dimensional system, but only the two and three dimensional systems
were explicitly solved. In this work we focus on the analytical evaluation
of the Canonical Partition Function (CPF) for two HS particles confined
in a Hard Wall spherical pore (HWSP) in arbitrary dimension $D$,
from now on 2-HS-HWSP. Then this work may be seen as the complement
and the dimensional generalization of \cite{Urru}. In the rest of
the manuscript we will use HS as the dimensional generalization of
Hard Spheres. With the purpose to avoid any confusion we should mention
that pore and cavity are synonyms along present work (PW). Sometimes,
an empty spherical space inside of a bulk fluid was also called a
cavity in the literature. 

From a more general point of view, the object of this manuscript is
the exact analytic study of an inhomogeneous \textit{fluid} system
with spherical interface. This general problem is currently studied
because of an incomplete understanding of the surface tension behaviour
in presence of curved interfaces \cite{Blok1,Blok3,Bryk,Mart,Row2}.
It seems that the spherical symmetry is the simplest one, and then
the principal subject of several works on curved interfaces but deviations
from sphericity are also studied \cite{Mart}. The suspended drop
on its vapor \cite{Blok1,Blok2,Hend}, the bubble of vapor on its
liquid, the fluid in contact with a spherical convex substrates (or
cavity in the liquid) \cite{Blok3,Bryk,Hend,Hend2}, and the fluid
confined in a spherical vessel or pore \cite{Blok3,HendD,Poni} are
different systems in which the spherical inhomogeneity of the fluid
is central and currently, the study of these systems are converging
to the analysis of the curvature dependence of physical magnitudes
\cite{Blok3}. Particularly relevant for PW are such works on HS systems
\cite{Bryk} and Hard Wall spherical substrates \cite{Blok3,Hend2}.
PW seen on this context, shows the dimensional dependence of an analytical
solvable system on this up to date and relevant problem in statistical
mechanics and thermodynamics. In PW we study a \textit{fluid} in contact
with a hard wall, therefore, we deal with the surface tension of a
fluid-substrate interface. We know that the point of view of a two-particle
\textit{fluid} may be somewhat conflicting. Few-body systems are not
gases, nor liquids and neither solids, but we wish to emphasize that
the toolbox of statistical mechanics must be applicable also to the
two particles inhomogeneous system, despite of whether it is fluid
or not. Naturally, the ergodic characteristics of the system must
be considered. We may note that the equivalence between different
ensembles of statistical mechanics is here invalid. The use of the
canonical ensemble enable the analysis of such few-body system. In
PW it is assumed that the 2-HS-HWSP system is well described by the
constant temperature ensemble without angular momentum conservation,
i.e. the usual canonical ensemble, but different assumptions may be
done \cite{HS_2inSph_micro}. The macroscopic open inhomogeneous system
of many HS, interacting with a Hard Wall and even in a HWSP were studied
in a virial series way by Bellemans \cite{Bell} and, Stecki and Sokolowski
\cite{Soko} for $D=3$. The virial series or density and curvature
power series expansion of statistical mechanic magnitudes are of the
highest importance, because are a source of exact results, which guide
the development of the field. Therefore, we will make contact between
PW and several cluster integrals reported in \cite{Bell,Soko}. The
first terms of the density and curvature expansion of the grand canonical
potential, surface tension, and adsorption are easily obtained as
a by product of the Configuration Integral (CI) of 2-HS-HWSP. Then,
we present the value of some integral coefficients related to thermodynamical
properties of open inhomogeneous systems and comment on this until
now unknown dimensional dependence.

In Section \ref{sec:Cap1} we show how a spherical pore that contains
two HS can be treated as another particle. There we bring up the central
problem solved in PW, the evaluation of CPF and distribution functions
of 2-HS-HWSP in $D$ dimensions. We also establish a relation between
its CI, the CI of three HS and the third pressure virial coefficient
for the bidisperse homogeneous system of HSs. The exact analytical
evaluation of its statistical mechanic properties is done in Section
\ref{sec:The-CI-integ} where we do a detailed inspection of the highest
and lowest density limits and make the link with the first terms of
the density power series coefficients of some physical properties
of the bulk fluid system. Principal characteristics of the one body
distribution function are also analyzed. The Section \ref{sec:Other-related-problems}
is devoted to trace the properties of two systems closely related
to the 2-HS-HWSP: two sticked spheres in a spherical pore and two
spheres moving on a spherical surface. The CI of the three systems
are strongly related as a consequence of the introduced sticky bond
transformation. A discussion of the mechanical equilibrium condition
for spherical inhomogeneous systems and the analytical evaluation
of the pressure tensor is reported in Section \ref{sec:Pressure-ten}.
The Equation of State, the (substrate-fluid) surface tension and other
thermodynamic features of the 2-HS-HWSP are studied in Section \ref{sec:EOS},
where the density and curvature first order terms of several magnitudes
are determined. Several relations with the bulk HS-HWSP open system
and its conjugated system, when HS particles are outside of a hard
spherical substrate, are also provided. Final remarks are given in
Section \ref{sec:Conclusions}.

\section{Partition function and Diagrams \label{sec:Cap1}}

We are interested in the properties of the CPF of two $D$-Spherical
particles of diameter $\sigma$, which are able to move inside of
a $D$-spherical pore of radius $R'$. The total partition function
$\tilde{Q}_{D}$ splits on kinetic and configuration space terms.
The kinetic term is given by $\Lambda^{-2D}$ where $\Lambda=(2\pi\beta\hbar^{2}/m)^{1/2}$
is the thermal de Broglie wavelength and $\beta=(k_{B}T)^{-1}$. Then
the central problem solved in this work corresponds to the evaluation
of CI, i.e., to analytically solve $Q_{D}$\begin{equation}
Q_{D}=\tilde{Q}_{D}\Lambda^{2D}=\int\int\, e_{\mathrm{A}}e_{\mathrm{B}}e_{\mathrm{AB}}\, d\mathbf{r}_{\mathrm{A}}d\mathbf{r}_{\mathrm{B}}\;,\label{eq:QD}\end{equation}
\begin{equation}
e_{\mathrm{AB}}=exp[-\beta\phi(r_{\mathrm{AB}})]=\Theta(r_{\mathrm{AB}}-\sigma)\;,\label{eq:e12}\end{equation}
\begin{equation}
e_{i}=exp[-\beta\psi(r_{i})]=\Theta(R-r_{i})\;,\label{eq:ei}\end{equation}
where $\phi(r_{\mathrm{AB}})$ is the hard core potential between
both spherical particles, $\psi(r_{i})$ is the external hard spherical
potential, $i=\mathrm{A},\mathrm{B}$ and $Q_{D}$ is independent
of temperature. The effective radius is $R=R'-\sigma/2$ and $\Theta(x)$
is the Heaviside unit step function ($\Theta(x)=1$ if $x\geq0$ and
zero otherwise). Then a variation of $\sigma$ at fixed $R$ does
not imply a volume variation. For future reference we introduce the
Mayer functions $f$ (or $f$-bond)\begin{eqnarray}
e_{\mathrm{AB}}=1+f_{\mathrm{AB}}\;, & \quad & f_{\mathrm{AB}}=-\Theta(\sigma-r_{\mathrm{AB}})\:,\nonumber \\
\nonumber \\e_{i}=1+f_{i}\;, & \quad & f_{i}=-\Theta(r_{i}-\sigma)\:,\label{eq:ef12i}\end{eqnarray}
functions $e_{\mathrm{A}}$, $e_{\mathrm{B}}$ and $f_{\mathrm{AB}}$
may be seen as overlapping functions, they take the unit value (positive
or negative) if certain pair of spheres overlaps and become null if
the pair of spheres does not overlap, just the opposite apply to the
$f_{\mathrm{A}}$, $f_{\mathrm{B}}$ and $e_{\mathrm{AB}}$ non overlapping
functions. We are interested in the graph representation of Eq. (\ref{eq:QD})
then we will draw the positive overlap functions, i.e. $\{e_{\mathrm{A}},e_{\mathrm{B}},-f_{\mathrm{AB}}\}$,
as continuous lines, and positive non overlap functions, i.e. $\{-f_{\mathrm{A}},-f_{\mathrm{B}},e_{\mathrm{AB}}\}$
as dashed lines. The graph of CI is then\begin{equation}
Q_{D}=\raisebox{-20pt}{\psfig{figure=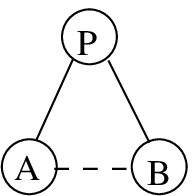,width=1.3cm}}\;,\label{eq:QDg}\end{equation}
where it is implicitly assumed the integration over the coordinates
of both particles $\mathrm{A}$ and $\mathrm{B}$. The Eq. (\ref{eq:QDg})
shows that the system is equivalent to a three particle system. We
can apply the in-out relation for three bodies \cite{Urru} performing
a simple decomposition over the particle $P$ in Eq. (\ref{eq:QDg})\begin{equation}
\raisebox{-20pt}{\psfig{figure=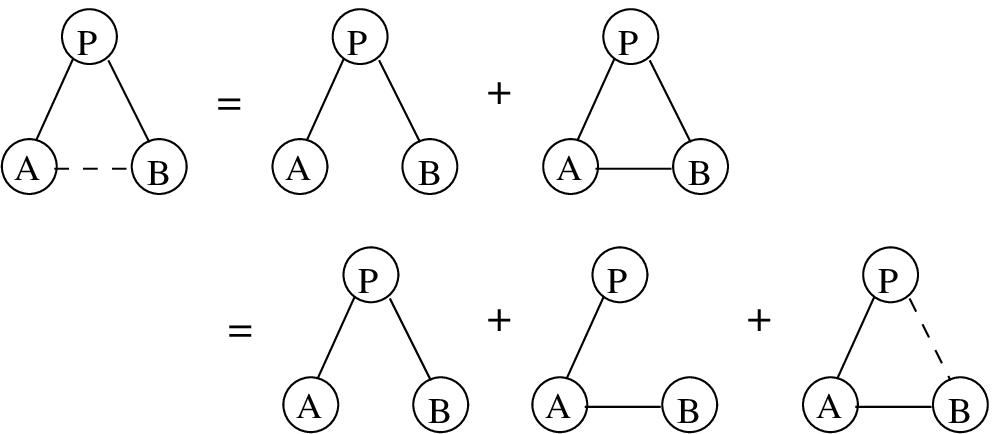,width=6.0cm}}\;.\label{eq:tri1}\end{equation}
where minus one factors were omitted. All the integrals drawn as not
fully connected graphs can be evaluated directly because they are
factorisable \cite{BookHill}. Focusing on the fully connected graphs,
first row relates the configuration integral with the -shape and volume
dependent- second cluster integral $b_{D}(V)$ of the inhomogeneous
system, which is also part of the third cluster integral of the homogeneous
multidisperse HS system \cite{BookHill}. Second row separates the
trivial volumetric term from the non trivial area-scaling integral.
For only two bodies we have\begin{equation}
\raisebox{-1pt}{\psfig{figure=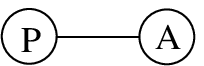,width=1.3cm}}=V_{D}(R)=R^{D}S_{D}/D\:,\label{eq:Vol}\end{equation}
\begin{equation}
\raisebox{-1pt}{\psfig{figure=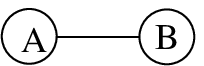,width=1.3cm}}=-V_{D}(\sigma)=-2b_{D}\:,\label{eq:2bgraph}\end{equation}
\begin{equation}
S_{D}=2\pi^{D/2}/\Gamma(D/2)\:,\label{eq:Sangle}\end{equation}
where $V_{D}(R)$ and $A_{D}(R)=R^{D-1}S_{D}$ are the volume and
surface area of the $D$-sphere of radius $R$, with $S_{D}$ the
solid angle. The Eq. (\ref{eq:Vol}) is the accessible volume for
a particle in a pore with effective radius $R$, i.e. the CI for the
one particle system, and Eq. (\ref{eq:2bgraph}) is twice the second
cluster integral or second pressure virial coefficient in the infinitely
homogeneous system of HS which we name $b_{D}$ \cite{BookHill}.
With this prescription $b_{D}$ is a positive defined constant. An
interesting point is that an inner sphere with radius $R-\sigma/2$
exist, when this sphere had the same radius of the particles, i.e.
$\sigma/2$ or $R=\sigma$, last integral in first row of Eq. (\ref{eq:tri1})
is (minus) $2\beta_{2}^{vir}$ where $\beta_{2}^{vir}$ is the second
irreducible cluster integral and $-2\beta_{2}^{vir}/3$ is the third
virial series coefficient of the pressure for the homogeneous HS system
\cite{BookHill} in $D$ dimensions.

\section{The density distribution and CI integration\label{sec:The-CI-integ}}

With the purpose of evaluate analytically $Q_{D}$ we introduce a
few geometrical functions. The function $W_{D}(r,R_{1})$, for $0\leq r\leq2R_{1}$,
is the partial overlap volume between two spheres of radius $R_{1}>0$
which centers are separated by a distance $r$, while for $-2R_{1}\leq r<0$,
it measures the joined volume of partial overlapping spheres. The
volumes of intersection and union of both spheres are related by $W_{D}(r,R_{1})=2V_{D}(R_{1})-W_{D}(-r,R_{1})$.
The function $Z_{D}(r,R_{1})$, is a generalization of the above idea
for $-\infty<r<+\infty$,\begin{equation}
\begin{array}{ccc}
Z_{D}(r,R_{1}) & = & \left\{ \begin{array}{ll}
0\:, & \mathrm{if}\: r>2R_{1}\:,\\
W_{D}(r,R_{1})\,, & \mathrm{if}\:\left|r\right|\leq2R_{1}\:,\\
2V_{D}(R_{1})\:, & \mathrm{if}\: r<-2R_{1}\:.\end{array}\right.\end{array}\label{eq:ZD}\end{equation}
Similarly, for two spheres with different radii $R_{1}$ and $R_{2}$
(assuming $R_{1}\geq R_{2}$) and restricting now to $r>0$ we have
the overlap volume $Z_{D}(r,R_{1},R_{2}),$\begin{equation}
\begin{array}{ccc}
Z_{D}(r,R_{1},R_{2}) & = & \left\{ \begin{array}{ll}
0\:, & \mathrm{if}\: r>R_{1}+R_{2}\:,\\
W_{D}(r,R_{1},R_{2})\,, & \mathrm{if}\: R_{1}-R_{2}\leq r\leq R_{1}+R_{2}\:,\\
V_{D}(R_{2})\:, & \mathrm{if}\:0<r<R_{1}-R_{2}\:,\end{array}\right.\end{array}\label{eq:ZDb}\end{equation}
where $W_{D}(r,R_{1},R_{2})$ is the volume in the partial overlap
configuration. As a consequence of the lens shape of the intersecting
volumes of two equal and unequal spheres, they are related by\begin{equation}
W_{D}(r,R_{1},R_{2})=\frac{1}{2}W_{D}(r',R_{1})+\frac{1}{2}W_{D}(r'',R_{2})\:,\label{eq:WDb}\end{equation}
with $r'=r+(R_{1}^{2}-R_{2}^{2})/r$, $r''=r-(R_{1}^{2}-R_{2}^{2})/r$
and negative values of $r'$ and $r''$ are possible due to Eq. (\ref{eq:ZD}).
We may transform Eq. (\ref{eq:ZD}) to a dimensionless function of
$x=r/(2R_{1})$\begin{equation}
Z_{D}(r,R_{1})=V_{D}(R_{1})\,\zeta_{D}(x)=V_{D}(R_{1})\times\left\{ \begin{array}{ll}
0\:, & \mathrm{if}\:\: x>1\:,\\
w_{D}(x)\,, & \mathrm{if}\:-1\leq x\leq1\:,\\
2\:, & \mathrm{if}\:\: x<-1\:,\end{array}\right.\label{eq:ZDz}\end{equation}
Function $w_{D}(x)$ for $0\leq x\leq1$, measures the overlap or
intersection volume between two spheres with unit radii separated
by a distance $x$, whereas $w_{D}(-x)$ measures the join volume.
Following the same idea Eq. (\ref{eq:ZDb}) may be expressed as a
function of $y=r/(2\bar{R})$ (with $\bar{R}=(R_{1}+R_{2})/2$) \begin{equation}
Z_{D}(r,R_{1},R_{2})=V_{D}(\bar{R})\,\zeta_{D}(y,\Delta)=V_{D}(\bar{R})\times\left\{ \begin{array}{ll}
0\:, & \mathrm{if}\:\: y>1\:,\\
w_{D}(y,\Delta)\,, & \mathrm{if}\:\Delta\leq y\leq1\:,\\
w_{D}(\Delta,\Delta)\:, & \mathrm{if}\:\:0\leq y<\Delta\:,\end{array}\right.\label{eq:ZDbzb}\end{equation}
and $\Delta=(R_{1}-R_{2})/(2\bar{R})$. For $\Delta=0$ one gets $y=x$,
$\zeta_{D}(y,0)=\zeta_{D}(x)$ and $w_{D}(y,0)=w_{D}(x)$. Function
$w_{D}(y,\Delta)$ measures the normalized overlap volume of two spheres
with unit mean radius, asymmetry $\Delta$, and centers separated
by a distance $y$ with $\Delta\leq y\leq1$. The analysis of the
properties of functions $\{\zeta_{D}(x),\zeta_{D}(y,\Delta),w_{D}(x),w_{D}(y,\Delta)\}$
will be left to next subsection \ref{sub:The-auxiliar-functions}.
Now we shall point out that, following a procedure depicted in \cite{Urru},
we may write down two relevant distribution functions. A pair distribution
function $\tilde{g}(r_{\mathrm{AB}})$ in which the position of the
pore center was integrated, and the one body distribution $\rho(r_{\mathrm{A}})$
\cite{BookHill}\begin{equation}
\tilde{g}(r_{\mathrm{AB}})=Q_{D}^{-1}V_{D}(R)\, e_{\mathrm{AB}}\, w_{D}(r_{\mathrm{AB}}/2R)\:,\label{eq:g2}\end{equation}
\begin{equation}
\rho(r_{\mathrm{A}})=2Q_{D}^{-1}V_{D}(R)\, e_{\mathrm{A}}\,\left[1-(\frac{R+\sigma}{2R})^{D}\,\zeta_{D}\left(\frac{r_{\mathrm{A}}}{R+\sigma},\frac{R-\sigma}{R+\sigma}\right)\right]\:.\label{eq:rho1}\end{equation}
Function $\tilde{g}(r_{\mathrm{AB}})$ is the probability density
distribution of finding both particles separated by a vector $\mathbf{r}_{\mathrm{AB}}$.
The normalization equations for these functions are $\int\rho(r_{\mathrm{A}})d\mathbf{r}_{\mathrm{A}}=2$
and $\int\tilde{g}(r_{\mathrm{AB}})d\mathbf{r}_{\mathrm{AB}}=1$.
We also introduce $g(r_{\mathrm{AB}})=\tilde{g}(r_{\mathrm{AB}})\, V_{D}(R)/2$
which is essentially the usual prescription of the pair distribution
function in the canonical ensemble (see Eq. (29.35) in \cite{BookHill}).
Performing the complete integration (of Eq. (\ref{eq:g2}) for example),
it is found the partition function\begin{eqnarray}
\tilde{Q}_{D} & = & \Lambda^{-2D}S_{D}V_{D}(R)\,\int_{\sigma}^{2R}\, r^{D-1}\, w_{D}(r/2R)\, dr\:,\label{eq:QD_1}\\
 & = & \Lambda^{-2D}V_{D}^{2}(R)\,2^{D}D\,\int_{z}^{1}\, x^{D-1}\, w_{D}(x)\, dx=\Lambda^{-2D}V_{D}^{2}(R)\, q_{D}(z)\:,\label{eq:QD_2}\end{eqnarray}
where $z=\sigma/(2R)$ is positive, $q_{D}(z)$ is the reduced CI,
and $q_{D}(z)=0$ if $z\geq1$.

\subsection{The auxiliary functions $w_{D}(x)$ and $w_{D}(y,\Delta)$\label{sub:The-auxiliar-functions}}

The overlap volume between two spheres with unit radii in any dimension,
$w_{D}(x)$ for $0\leq x\leq1$, introduced in Eq. (\ref{eq:ZDz})
is \cite{Baus}\begin{equation}
w_{D}(x)=I_{1-x^{2}}((D+1)/2,1/2)\:,\label{eq:wDI}\end{equation}
being $I_{x}(a,b)=B_{x}(a,b)/B(a,b)$ the normalized incomplete beta
function, defined in terms of the beta function $B(a,b)=\Gamma(a)\Gamma(b)/\Gamma(a+b)$
and the incomplete beta function $B_{x}(a,b)=\intop_{0}^{x}dt\, t^{a-1}(1-t)^{b-1}$
\cite{BookAbram}. For future reference we introduce the shortcut
$\mathtt{B}_{D}\equiv1/B((D+1)/2,1/2)=S_{D+1}/S_{D+2}$. We extend
definition (\ref{eq:wDI}) following Eqs. (\ref{eq:ZD}, \ref{eq:ZDz})
to $w_{D}(x)-1=-(w_{D}(-x)-1)$ for $-1\leq x\leq0$. In turn, due
to the relation $I_{1-x^{2}}(a,b)=1-I_{x^{2}}(b,a)$, for $-1\leq x\leq1$
we have\begin{eqnarray}
w_{D}(x)-1 & \equiv- & sign(x)\, I_{x^{2}}(1/2,(D+1)/2)\:,\label{eq:weDI}\end{eqnarray}
where $sign(x)=1$ if $x\geq0$ and $sign(x)=-1$ if $x<0$. Therefore,
for $D>0$, $w_{D}(x)$ is analytic for $-1<x<1$ and both $\{w_{D}(x)-1,\:\zeta_{D}(x)-1\}$
are odd functions. It is also possible to construct $w_{D}(x)$ from
the recurrence relations \cite{Baus}

\begin{equation}
w_{-1}(x)=1\:,\label{eq:wm1}\end{equation}
\begin{equation}
w_{0}(x)=\frac{2}{\pi}\,\arccos(x)\:,\label{eq:w0}\end{equation}
\begin{equation}
w_{D}(x)=w_{D-2}(x)-x\left(1-x^{2}\right)^{(D-1)/2}2\mathtt{B}_{D}/D\:,\label{eq:wDrec}\end{equation}
 $\mathtt{B}_{-1}=0$, $\mathtt{B}_{0}=1/\pi$ and $\mathtt{B}_{D}=D/(2\pi\mathtt{B}_{D-1})$.
Expressions for $D=1,\,2,\,3,\,4,$ and $5$ are\begin{equation}
w_{1}(x)=1-x\:,\label{eq:w1}\end{equation}
\begin{equation}
w_{2}(x)=\frac{2}{\pi}\,\left[\arccos(x)-x\left(1-x^{2}\right)^{1/2}\right]\:,\label{eq:w2}\end{equation}
\begin{equation}
w_{3}(x)=1-\frac{3}{2}x+\frac{1}{2}x^{3}\:,\label{eq:w3}\end{equation}
\begin{equation}
w_{4}(x)=\frac{2}{\pi}\,\left[\arccos(x)-\left(\frac{5}{3}x-\frac{2}{3}x^{3}\right)\left(1-x^{2}\right)^{1/2}\right]\:,\label{eq:w4}\end{equation}
\begin{equation}
w_{5}(x)=1-\frac{15}{8}x+\frac{5}{4}x^{3}+\frac{3}{8}x^{5}\:.\label{eq:w5}\end{equation}
As far as we will need the asymptotic analysis of the $w_{D}(x)$
function, it is resumed here. In the $x\rightarrow0$ and $x\rightarrow1$
limits we have respectively 

\begin{equation}
w_{D}(x)=1-2\,\mathtt{B}_{D}\, x+\frac{D-1}{3}\mathtt{B}_{D}\, x^{3}-O_{5}(x)\:,\label{eq:wDser1}\end{equation}
\begin{equation}
w_{D}(x)=\frac{2^{(D+3)/2}\mathtt{B}_{D}}{D+1}\,(1-x)^{(D+1)/2}\,\left(1+O(1-x)\right)\:.\label{eq:wDser2}\end{equation}

Following Eqs. (\ref{eq:ZD})-(\ref{eq:ZDz}) $w_{D}(y,\Delta)$ may
be written as\begin{equation}
w_{D}(y,\Delta)=\frac{1}{2}\left(1+\Delta\right)^{D}w_{D}(x')+\frac{1}{2}\left(1-\Delta\right)^{D}w_{D}(x'')\:.\label{eq:wyDdef}\end{equation}
with $x'=r'/2R_{1}=\cos(\theta_{1})=(y^{2}+\Delta)/[y\,(1+\Delta)]$
and $x''=r''/2R_{2}=\cos(\theta_{2})=(y^{2}-\Delta)/[y\,(1-\Delta)]$,
here $\theta_{i}$ is the angle opposite to the side of length $R_{i}$
in the triangle $(R_{1},R_{2},r)$, i.e. $\cos(\theta_{1})=(-R_{1}^{2}+R_{2}^{2}+r^{2})/(2R_{2}r)$
and $\cos(\theta_{2})=(R_{1}^{2}-R_{2}^{2}+r^{2})/(2R_{1}r)$. A recurrence
relation for $w_{D}(y,\Delta)$ is derived in the Appendix \ref{appendix-A}.
From Eqs. (\ref{eq:wm1})-(\ref{eq:w5}) or from Eqs. (\ref{eqa:wyDrec})-(\ref{eqa:wDym})
of the Appendix \ref{appendix-A} we have that the first functions
of the series are\begin{equation}
w_{-1}(y,\Delta)=1/(1-\Delta^{2})\:,\label{eq:wym1}\end{equation}
\begin{equation}
w_{0}(y,\Delta)=v_{0}(y,\Delta)\:,\label{eq:wy0}\end{equation}
\begin{equation}
v_{D}(y,\Delta)=\frac{1}{\pi}(1+\Delta)^{D}\arccos(\frac{y^{2}+\Delta}{y\,(1+\Delta)})+\frac{1}{\pi}(1-\Delta)^{D}\arccos(\frac{y^{2}-\Delta}{y\,(1-\Delta)})\:,\label{eq:arf}\end{equation}
leading to the following expressions for $D=1,\,2,\,3,\,4,$ and $5$
\begin{equation}
w_{1}(y,\Delta)=1-y\:,\label{eq:wy1}\end{equation}
\begin{equation}
w_{2}(y,\Delta)=-\frac{2}{\pi}\,\left((1-y^{2})(y^{2}-\Delta^{2})\right)^{1/2}+v_{2}(y,\Delta)\:,\label{eq:wy2}\end{equation}
\begin{equation}
w_{3}(y,\Delta)=(2y)^{-1}(1-y)^{2}(2y+y^{2}-3\Delta^{2})\:,\label{eq:wy3}\end{equation}
\begin{equation}
w_{4}(y,\Delta)=-\frac{2}{3\pi}\, y^{-2}\left((1-y^{2})(y^{2}-\Delta^{2})\right)^{1/2}(2y^{4}-5y^{2}(1+\Delta^{2})-4\Delta^{2})+v_{4}(y,\Delta)\:,\label{eq:wy4}\end{equation}
\begin{equation}
w_{5}(y,\Delta)=(2y)^{-3}(1-y)^{3}(8y^{3}+9y^{4}+3y^{5}-10\Delta^{2}y^{2}(3+y)+5\Delta^{4}(1+3y))\:,\label{eq:wy5}\end{equation}
where Eqs. (\ref{eq:wym1})-(\ref{eq:wy5}) apply for $max(0,\Delta)\leq y\leq1$.

\subsection{The reduced CI}

\begin{figure}
\begin{centering}
\includegraphics[clip,width=7cm]{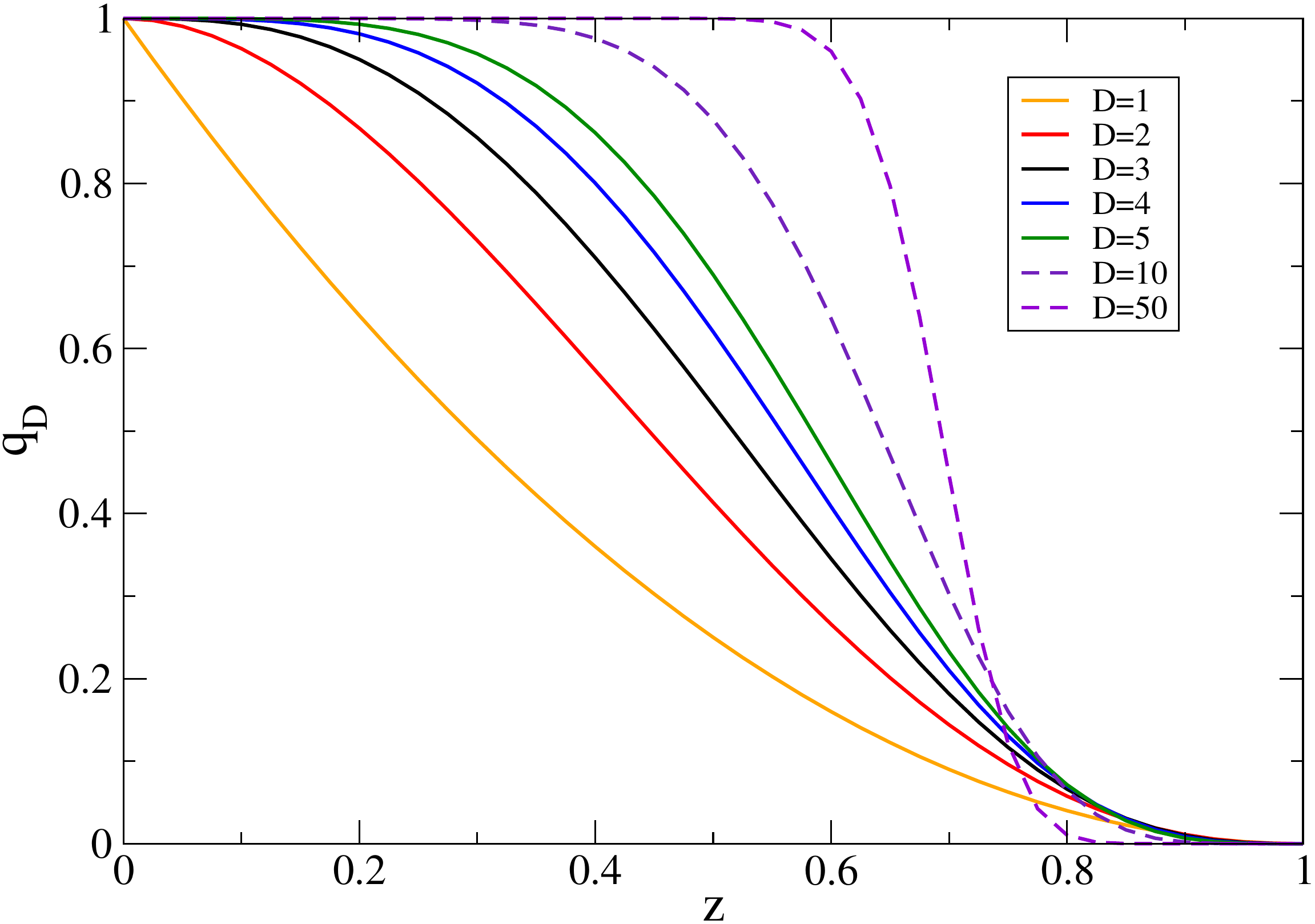}
\par\end{centering}

\caption{(color online) Reduced partition function $q_{D}(z)$ as a function
of $z=\sigma/2R$ for several dimensions. From left to right $D=1,\,2,\,3,\,4,5,10,$
and $50$ last two in dashed line.\label{fig:Adimentional-partition-function}}

\end{figure}

As shown in the Appendix \ref{appendix-B}, the reduced CI $q_{D}(z)$
defined in Eq. (\ref{eq:QD_1}) may be written as\begin{equation}
q_{D}(z)=u_{D}(z)-(2z)^{D}w_{D}(z)\:,\label{eq:quw}\end{equation}
 for $0\leq z\leq1$ while $q_{D}(z>1)=q_{D}(1)=0$, with\begin{equation}
u_{D}(z)=I_{1-z^{2}}((D+1)/2,(D+1)/2)\;,\label{eq:uDef}\end{equation}
valid for $D\geq0$ but not for $D=-1$. Quantity $u_{D}$ may be
derived from the recurrence relation\begin{equation}
u_{-1}(z)=1/2\:,\label{eq:um1}\end{equation}
\begin{equation}
u_{0}(z)=(2/\pi)\,\arccos(z)\:,\label{eq:u0}\end{equation}
\begin{equation}
u_{D}(z)=u_{D-2}(z)+z^{D-1}(1-z^{2})^{(D-1)/2}(\frac{1}{2}-z^{2})\Gamma(D)/\Gamma^{2}((D+1)/2)\:.\label{eq:uiter}\end{equation}
The CI for $D=-1$ and $0$ are\begin{equation}
q_{-1}(z)=\frac{1}{2}(1-\frac{1}{z})\:,\label{eq:qm1}\end{equation}
\begin{equation}
q_{0}(z)=0\:,\label{eq:q0}\end{equation}
while for $D=1,\,2,\,3,\,4,$ and $5$ are

\begin{equation}
q_{1}(z)=\left(1-z\right)^{2}\:,\label{eq:q1}\end{equation}

\begin{equation}
q_{2}(z)=\frac{2}{\pi}\left[\left(1-(2z)^{2}\right)\arccos(z)+z\,\sqrt{1-z^{2}}\left(1+2z^{2}\right)\right]\:,\label{eq:q2}\end{equation}
\begin{equation}
q_{3}(z)=1-(2z)^{3}+9z^{4}-2z^{6}\:,\label{eq:q3}\end{equation}
\begin{equation}
q_{4}(z)=\frac{2}{\pi}\left[\left(1-(2z)^{4}\right)\arccos(z)+\frac{1}{3}z\,\sqrt{1-z^{2}}\left(3+2z^{2}+56z^{4}-16z^{6}\right)\right]\:,\label{eq:q4}\end{equation}
\begin{equation}
q_{5}(z)=1-(2z)^{5}+50z^{6}-25z^{8}+6z^{10}\:.\label{eq:q5}\end{equation}
It is apparent that CI for odd $D$ is polynomial with order $2D$
and integer non null coefficients at terms of order $0,\, D$ and
$D+1+2k$ with $k=0,1,2,...,(D-1)/2$. However, for even $D$ partition
function is not polynomial. These and other interesting properties
may be derived from the series representation of the incomplete beta
function (from \cite{Functions} and Eq. (26.5.4) of \cite{BookAbram}).
Function $q_{D}(z)$ is plotted in Fig. \ref{fig:Adimentional-partition-function}
for several values of $D$. For all $D\geq0$ we have $q_{D}(0)=1$,
$q_{D}(1)=0$, and for large $D$ limit $q_{D\rightarrow\infty}(z)=\Theta(2^{-1/2}-z)$.
Let us now look at some asymptotic behavior. %

\paragraph{Large cavity limit:}

The infinitely dilution or large cavity limit $z\rightarrow0$ (i.e.
$\sigma/R\rightarrow0$) can be studied from the series representation
of the reduced partition function $q_{D}(z)$ near $z\simeq0$,

\begin{equation}
q_{D}(z)=1-z^{D}C_{1,0}+z^{D+1}C_{1,1}+z^{D+3}C_{1,2}+\sum_{k=1}^{\infty}z^{D+3+2k}C_{1,2+k}\;,\label{eq:qDser1}\end{equation}
where we introduce a set of dimensional dependent constants $\{C\}$
that will denote some power series coefficients through PW. Here,
$C_{1,0}=2^{D}$, $C_{1,1}=2^{D}\mathtt{B}_{D}\frac{2D}{D+1}$ and
$C_{1,2}=-2^{D}\mathtt{B}_{D}\frac{D(D-1)}{3(D+3)}$. For odd $D$
Eq. (\ref{eq:qDser1}) is effectively an order $2D$ polynomial. As
we will see later, the constant coefficients $\{C_{1,0},\, C_{1,1},\, C_{1,2}\}$
are involved in physical properties of the equivalent many body system.
From Eqs. (\ref{eq:QD}) and (\ref{eq:QD_2}) we obtain \begin{equation}
Q_{D}=V_{D}^{2}(R)-V_{D}(R)\,2b_{D}+A_{D}(R)\,2a_{D}-J\!\! J_{D}(R)\,2\delta_{D}^{(1)}+O_{D-5}(R)\;,\label{eq:QDser1}\end{equation}
written in terms of the extensive squared mean curvature $J\!\! J_{D}(R)=A_{D}(R)\cdot(D-1)^{2}\cdot R^{-2}$.
The coefficients of area and curvature are \begin{equation}
a_{D}=b_{D+1}(2\pi)^{-1}\:,\label{eq:aD}\end{equation}
\begin{equation}
\delta_{D}^{(1)}=b_{D+3}\frac{D+1}{D-1}\,(2^{5}3\pi^{2})^{-1}\:,\label{eq:delta1D}\end{equation}
and $\delta_{D}^{(0)}=0$ corresponds to the absent term proportional
to $A_{D}(R)\, R^{-1}$. We may remark that curvature dependence is
neither proportional to total nor to Gaussian curvatures, $\mathrm{j}=(D-1)/R$
and $\mathrm{k}=R^{-(D-1)}$ respectively. As may be expected $V_{D}(R)\,2b_{D}$
is the first non ideal correction, and first sign of inhomogeneity
and curvature dependence appears in the next two terms. They are deeply
connected with $b_{D+j}$ the second cluster integral in higher dimensionality.
Therefore, we are showing a direct relation between the -intrinsically
inhomogeneous- properties of 2-HS-HWSP and the low density limit properties
of HS homogeneous system in higher dimension. Equations (\ref{eq:QDser1},
\ref{eq:aD}) and (\ref{eq:delta1D}) also mixes the properties of
systems with odd and even dimensions. Finally, note that Eq. (\ref{eq:QDser1})
is exact for $D=3$ without the order $D-5$ term \cite{Urru} and
coincidentally $\mathrm{j}^{2}\sim\mathrm{k}$ with the constant value
$J\!\! J_{3}(R)=2^{4}\pi$. Further,  $J\!\! J_{D}(R)\,\frac{D+1}{D-1}|_{D=3}=J\!\! J_{3}(R)+2K_{3}(R)$
with the extensive gaussian curvature $K_{D}(R)=A_{D}(R)\cdot\mathrm{k}$.
Fixing $R=\sigma$ the inner sphere and both particles have the same
size and we obtain from Eqs. (\ref{eq:tri1}, \ref{eq:QDser1})\begin{equation}
2\beta_{2}^{vir}\mid_{D=3}=-(2b_{3})^{2}+A_{3}(\sigma)\,2a_{3}-J\!\! J_{3}(\sigma)\,2\delta_{3}^{(1)}\;,\label{eq:betavir}\end{equation}
where the second irreducible cluster integral $\beta_{2}^{vir}$ which
involves three nodes \cite{BookHill}, is now written in terms of
geometrical measurements of the cavity's boundary $\{A_{3}(\sigma),\, J\!\! J_{3}(\sigma)\}$
in three dimensions and two body integrals $b_{D}$ in $D\geq3$.
Baus and Colot \cite{Baus} found $\beta_{2}^{vir}\mid_{D}=-(b_{D})^{2}w_{D}(1/2)$.
These type of relations may be interesting for higher order integrals.
The three body integral (two HS plus the cavity) $Q_{D}$ and its
moments are linked with several three body integrals describing physical
properties of the HS inhomogeneous fluid in the low density limit,
inside a spherical cavity and even outside a spherical substrate,
which will be discussed later in Sec. \ref{sec:EOS}.

\paragraph{Small cavity limit:}

Looking at the opposite situation of high density or caging limit
we found the \textit{final solid}, i.e., the densest available configuration
of the system. This limiting behaviour has been extensively studied
for the HS homogeneous solid system, but also, in small systems \cite{Nemeth,Urru}.
The caging limit of 2-HS-HWSP is obtained at the root $z=1$. The
reduced partition function has an interesting series representation
in the neighboring of $z=1$ (valid for $z\leq1$) \begin{equation}
q_{D}(z)=(1-z^{2})^{(D+3)/2}C_{2,0}\,(1+C_{2,1}\,(1-z^{2})+O_{2}(1-z^{2}))\;,\label{eq:qDser2}\end{equation}
with $C_{2,0}=C_{1,1}/(D+3)$, $C_{2,1}=\frac{7-D^{2}}{2(D+5)}$.
It shows that CI goes to zero as $(1-z)^{(D+3)/2}$ when system becomes
caged. The partition function is then \begin{equation}
Q_{D}\simeq\sigma^{3(D-1)/2}(R-\sigma/2)^{(D+3)/2}C_{3,0}(1-\sigma^{-1}C_{3,1}(R-\sigma/2)+O_{2}(1-2R/\sigma))\;,\label{eq:QDser2}\end{equation}
here $C_{3,0}=C_{2,0}\,2^{3-D}(S_{D}/D)^{2}$ and $C_{3,1}=2^{-1}(-9+8\, C_{2,1}+5D)$.
Since $Q_{D}$ is identically zero for $R<\sigma/2$ Eq. (\ref{eq:QDser2})
implies that $Q_{D}$ is non-analytic at $R=\sigma/2$. For odd $D$,
the derivative of order $2+(D+1)/2$ and beyond are zero. However,
for even $D$ the derivatives of order $2+D/2$ and bigger involves
an infinite discontinuity. When a closed system of hard particles
is caged as a consequence of the high confinement, the spatial degrees
of freedom that becomes lost are related to a zero measure set in
the CI integral similar to (\ref{eq:QD}). The consequence is that
$Q$ becomes zero, and the signature of the frozen spatial freedom
is the order of this root. We introduce the number of spatial lost
degrees of freedom $LDF$ and the complementary number of kept degrees
of freedom $KDF$, their addition provides the total spatial degrees
of freedom $LDF+KDF=N\cdot D$ where $N=2$ is the actual number of
particles. We relate $LDF$ with the exponent $(D+3)/2$ in (\ref{eq:QDser2}).
For the studied system of 2-HS-HWSP in the caging limit we obtain
\begin{equation}
LDF=(D+3)/2=1+(D+1)/2\:,\label{eq:ldf}\end{equation}
\begin{equation}
KDF=2D-LDF=(D-1)\,3/2\:.\label{eq:nldf}\end{equation}
In next sections we will study $LDF$ and $KDF$ in other situations.

\subsection{The one body distribution}

\begin{figure}
\centering{}\includegraphics[clip,width=7cm]{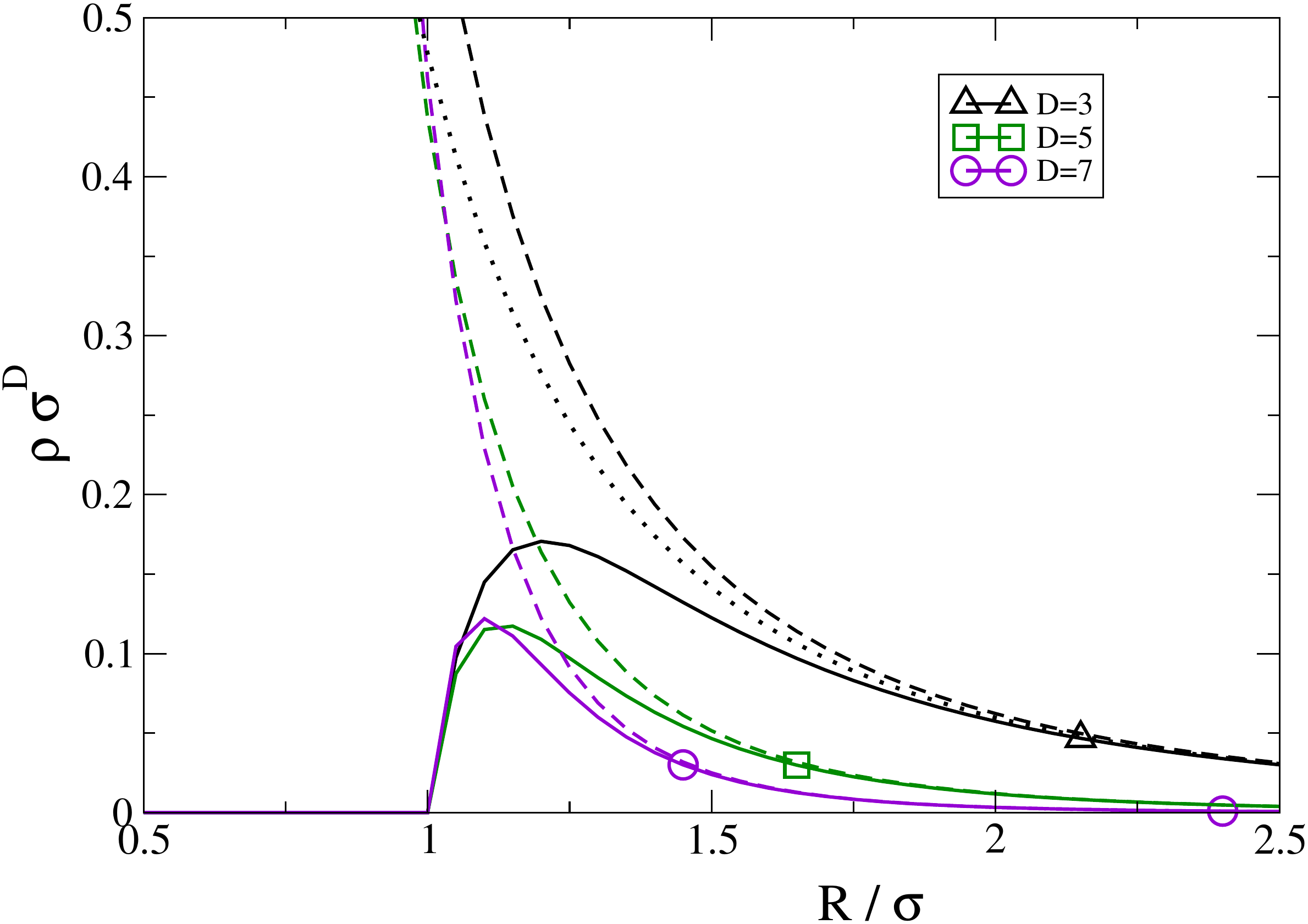}\caption{(color online) Characteristic values of density profile. Densities,
at the central plateau $\rho_{0}$ (continuous line) and at contact
$\rho_{c}$ (dashed line) as a function of pore size. From top to
bottom dimensions $D=3,\,5,$ and $7$. Rough density $\bar{\rho}$
for $D=3$ is showed in dotted line.\label{fig:density-min-max}}

\end{figure}

The one body distribution function are qualitatively similar for all
the dimensions and was previously described in detail for $D=2$ and
$3$ \cite{Urru} (Fig.2 therein). A peculiarity of $\rho(r)$ is
the plateau of constant density that appears for $R>\sigma$, and
which extends from the center to $r=R-\sigma$. For $\sigma/2<R<\sigma$
a null density plateau develops in the range $0<r<\sigma-R$. Therefore,
the central plateau density is\begin{equation}
\rho_{0}=2Q_{D}^{-1}V_{D}\,\left[1-(2z)^{D}\right]=2\left(V_{D}-2b_{D}\right)/Q_{D}\:,\label{eq:rho0}\end{equation}
if $R\geq\sigma$, while $\rho_{0}=0$ if $\sigma/2\leq R<\sigma$.
Other interesting magnitude is the density at the wall or contact
density $\rho_{c}=\rho(r=R)$,\begin{equation}
\rho_{c}=2V_{D}\, Q_{D}^{-1}\left\{ 1-\frac{1}{2}\left[w_{D}(1-2z^{2})+(2z)^{D}w_{D}(z)\right]\right\} \:.\label{eq:rhoc}\end{equation}
These quantities are shown for several dimensions as a function of
pore size in Fig. \ref{fig:density-min-max}. We introduce the rough
or mean number density $\bar{\rho}=2/V_{D}(R)$ for comparison. We
would like to emphasize that $\rho(r)$ is a discontinuous function
at the cavity surface falling down to zero outside the cavity and,
also, is non-analytic at $r=Abs(R-\sigma)$. The function $\varrho(r)$
a regularized version of $\rho(r)$ at $r=R$, may be introduced by
the short-hand \begin{equation}
\rho(r)=e_{A}(r)\,\varrho(r)\:,\label{eq:rhovar}\end{equation}
which must be understood with the help of Eq. (\ref{eq:rho1}). It
is similar to $\rho(r)$ but includes its analytic continuation for
$R\leq r\leq R+\sigma$ and therefore is a smooth function at the
hard wall (\cite{Hansen}, p.166 therein, this procedure was also
used in the case of homogeneous fluids, see Eq. (6.11) in \cite{Barker}).%

\section{Other Closely Related Systems\label{sec:Other-related-problems}}

It is possible to establish interesting relations between the CPF
of 2-HS-HWSP and that of other systems. One of them involves the one
stick or dumbbell in the spherical pore, where we think the stick
as a rigid body formed by two sticked HS. Other system is that composed
by two hard spherical bodies which are constrained to move between
the HWSP and an inner hard spherical core. For sufficiently large
hard core radius we find the limiting case of 2-HS that are able to
move on the surface of a sphere. We may imagine that this system is
constituted by two HS sticked to the surface of the HWSP, all immersed
in a $D$ dimensional euclidean space. Therefore, it has an effectively
reduced dimension of $D-1$. Both systems may be seen as originated
in the 2-HS-HWSP and they emerge as a consequence of the addition
of a new sticky property. Interestingly, this property is easily introduced
by a simple transformation. In PW we will not make a general digression
about this transformation, but we simple state that a \textit{sticky-bond
transformation} is performed when one or several of the $e$ or $f$
functions in (\ref{eq:QD}, \ref{eq:QD_2}) is transformed in a Dirac
delta function or $s$-bond. In a system of many hard bodies in a
hard pore a sticky transformation may be done for example by sticking
one particle on the surface of the pore, or sticking particles together.
The sticky-bond transformation is interesting because in an $N$-body
system the most compact spatial configuration, or final solid may
be obtained through a series of such operations. It may also mimic
a nucleation or condensation process, the adsorption on a surface
and even some chemical reactions. In this section the temperature
and kinetic factor are not relevant, then we simply make $\Lambda=1$
and then CI and CPF are equal. We may mention that the sticky bond
implies that the two bodies \textit{must} fix their distance, then,
even when the concept is related with the adhesive hard sphere bond
of Baxter \cite{Baxter} they are not equivalent.

\subsection{The one stick in a HWSP Partition Function\label{sec:The-rigid-body}}

When two HS become stickied together they conform a rigid body because
the $s$-bond fix the inter-particle distance to $\sigma$, they form
a stick. The sticky-bond transformation may be done by applying the
derivative $\partial_{\sigma}$ to $Q_{D}$ on Eqs. (\ref{eq:QD},
\ref{eq:QD_2}) \begin{equation}
Q_{D/b}=C_{b}\,\partial_{\sigma}Q_{D}\:,\label{eq:Qrig_0}\end{equation}
the $b$-subindex label means \textit{body-particle} opposite to \textit{point-particle}
which may be used to describe a HS. The constant $C_{b}=-\sigma^{-(D-1)}$
is found from the large size limit, obtaining \begin{equation}
Q_{D/b}=S_{D}\, Q_{D}\,\tilde{g}(\sigma)=S_{D}\, V_{D}(R)\, w_{D}(z)\:.\label{eq:Qrig_1}\end{equation}
Iterative construction of $w_{D}(z)$ and limits $z\rightarrow0$
and $z\rightarrow1$ were described in Section \ref{sub:The-auxiliar-functions}.
Limiting behaviour of the partition function $Q_{D/b}$ is, from Eq.
(\ref{eq:wDser1})\begin{eqnarray}
Q_{D/b} & = & S_{D}V_{D}(R)-A_{D}(R)\, S_{D+1}\sigma(2\pi)^{-1}-J\!\! J_{D}(R)\, S_{D+3}\frac{D+1}{D-1}\sigma^{3}(2^{5}3\pi^{2})^{-1}+O_{D-5}(R)\:,\label{eq:Qrigser1}\\
 & = & S_{D}\left(V_{D}(R)-A_{D}(R)\,\sigma\,\mathtt{B}_{D}/D-J\!\! J_{D}(R)\,\sigma^{3}\mathtt{B}_{D}\,(4!D(D-1))^{-1}\right)+O_{D-5}(R)\:,\label{eq:Qrigser2}\end{eqnarray}
Note that $V_{D}(R)$ is the volume for one HS particle in the spherical
cavity but $Vol=Q_{D/b}/S_{D}$ is the actual volume of the pore for
the rigid particle composed by two point particles with fix separation
$\sigma$, they are strongly different only for small $R$ in fact\begin{equation}
\frac{V_{D}(R)-Vol(R)}{V_{D}(R)}=\frac{\sigma}{R}\mathtt{B}_{D}\,\left[1-\left(\frac{\sigma}{R}\right)^{2}\frac{D-1}{4!}\right]+O_{5}(\frac{\sigma}{R})\:.\label{eq:Volstick}\end{equation}
The first correction to the one HS CPF is related to the Volume coefficient
of $Q_{D}$ and the next one relates to the Area coefficient as a
consequence of Eqs. (\ref{eq:QDser1}, \ref{eq:Qrig_1}). The relation
may be compactly written $\sigma^{D-1+m}S_{D+m}=2\partial_{\sigma}b_{D+m}$
and is a direct consequence of Eqs. (\ref{eq:Vol}, \ref{eq:2bgraph}).
The power at which $Q_{D/b}\rightarrow0$ .i.e $LDF$ is now, from
Eq. (\ref{eq:wDser2}), \begin{equation}
LDF=(D+1)/2\:.\label{eq:ldf02}\end{equation}
When this system becomes caged it conforms a linear rigid rotator.
The difference between $LDF$ in Eq. (\ref{eq:ldf02}) and $LDF$
in Eq. (\ref{eq:ldf}) is attributable to the relative distance between
both HS. Then we interpret that $LDF$ of $Q_{D}$ corresponds to
one degree of freedom from the relative distance between both HS and
$(D+1)/2$ corresponding to the $LDF$ of $Q_{D/b}$ related to the
rigid rotator degrees of freedom.

\subsection{The spherical pore with a hard core\label{sec:Pore_wCore}}

Other closely related system is that of two particle confined in a
spherical pore with a fixed central hard core. The main interest to
solve this pore shape is the study of the dimensional crossover between
$D$ and $D-1$ dimension, which happens when the internal core becomes
so large that the two HS may only move on the surface of the $D$-sphere,
an effective $D-1$ curved dimensional space. This type of non planar
surfaces was introduced by Kratky \cite{Kratky} in the study of the
HS homogeneous systems. Exact results for two, three and four hard
particles confined in this non euclidean surface embedded in $D=3$
have been reported \cite{calotes}. Besides, the 2-HS-HWSP with a
hard core (2-HS-HWSP+HC) is by itself interesting cause it is a highly
non trivial concave pore. The actual volume of the system is $V_{D}(R)-V_{D}(R-h)$,
being $R-h\geq0$ the radius of the core and $0\leq h<R$. The CI
of 2-HS-HWSP+HC was written in Eq. (\ref{eq:QD}) but Eq. (\ref{eq:ei})
must be replaced by $e_{i}=exp[-\beta\psi(r_{i})]=\Theta(R-r_{i})-\Theta(R-h-r_{i})$.
Expanding the product of Heaviside functions we obtain\begin{eqnarray}
Q_{D}(R,h) & = & S_{D}V_{D}(R)\,\int_{\sigma}^{2R}r^{D-1}w_{D}(r/2R)\, dr+S_{D}V_{D}(R-h)\,\int_{min(\sigma,2(R-h))}^{2(R-h)}r^{D-1}w_{D}(r/2(R-h))\, dr\nonumber \\
 &  & -2S_{D}V_{D}(R-h/2)\,\int_{min(\sigma,2R-h)}^{2R-h}r^{D-1}\zeta_{D}(r/(2R-h),h/(2R-h))\, dr\:,\label{eq:Q2P_def}\end{eqnarray}
where we have made explicit the conditions on the integration intervals.
From them, we observe that CI breaks in three branches depending on
$\sigma\leq2(R-h)$ or $2(R-h)<\sigma\leq2R-h$ or $2R-h<\sigma\leq2R$.
Last row, also shows that CI separates in two branches according to
Eq. (\ref{eq:ZDbzb}) depending on $\sigma\leq h$ or not. Equation
(\ref{eq:Q2P_def}) becomes \begin{eqnarray}
Q_{D}(R,h) & = & V_{D}^{2}(R)\, q_{D}(z)+V_{D}^{2}(R-h)\,\Theta(2(R-h)-\sigma)\, q_{D}(\sigma/2(R-h))\nonumber \\
 &  & -2V_{D}^{2}(\bar{R})\,\Theta(2R-h-\sigma)\,2^{D}D\,\int_{\tau}^{1}\, y^{D-1}\, w_{D}(y,\Delta)\, dy\label{eq:Q2P_def01}\\
 &  & -2\Theta(2R-h-\sigma)\Theta(h-\sigma)\, V_{D}(R-h)(V_{D}(h)-V_{D}(\sigma))\:,\nonumber \end{eqnarray}
being $\bar{R}=R-h/2$ and $\tau=max(\Delta,\sigma/2\bar{R})$. It
is clear from Eqs. (\ref{eq:Q2P_def01}, \ref{eq:wyDdef}) that at
each branch $Q_{D}$ is a finite power series in $R$, $h$ and $\sigma$
only for odd dimension. The dimensional systematic of the integral
in (\ref{eq:Q2P_def01}) becomes a complex task, then we only analyse
with more detail $D=3$ and some characteristics for $D=2,4,$ and
$5$. For $D=3$ we evaluate the integral to obtain an explicit expression
for the partition function, here we present the result for $\sigma\geq2(R-h)$,\begin{equation}
Q_{3}(R,h)=\begin{cases}
Vol^{2}-Vol\,2b_{3}+Ar\,2a_{3}-Cur\,2\delta_{3}^{(1)}\:, & \mathrm{if}\: h\geq\sigma\:,\\
\\Vol^{2}-Vol\,2b_{3}+Ar\,2a_{3}\lambda\,(8-6\lambda+\lambda^{3})/3+\\
-Cur\,2\delta_{3}^{(1)}\lambda^{3}\,(8-9\lambda+2\lambda^{3})\:, & \mathrm{if}\: h<\sigma\:,\end{cases}\label{eq:Qg01}\end{equation}
where $\lambda=h/\sigma$, and $Vol$, $Ar$ and $Cur$ are the measurements
of volume, surface area and boundary quadratic curvature for the actual
pore, i.e. $Vol=V_{3}(R)-V_{3}(R-h)$, $Ar=A_{3}(R)+A_{3}(R-h)$ and
$Cur=J\!\! J_{3}(R)+J\!\! J_{3}(R-h)=2^{5}\pi$. Noticeable, the partition
function is polynomial in each domain, and has continuous second derivative
in $h=\sigma$ but a discontinuous third derivative. It is surprising
that the volume coefficient $2b_{3}$ and the entire partition function
when $h\geq\sigma$ looks exactly equal to $Q_{3}$ (see Eq. (\ref{eq:QDser1})
and Ref. \cite{Urru}) with different Volume, Area, and Curvature
measures. We also have established that the central position of the
fixed hard core is not essential, first row of Eq. (\ref{eq:Qg01})
is valid for any internal fixed hard core as long as its boundary
is separated from the outer spherical pore wall a distance $\geq\sigma$.
The limiting behaviour as $h\rightarrow0$ to first non null order
in $h$ is, from Eq. (\ref{eq:Qg01})\begin{equation}
Q_{3}(R,h\ll\sigma)=h^{2}\left(A_{3}^{2}(R)-A_{3}(R)\,\pi\sigma^{2}\right)+O_{3}(h)=h^{2}S_{3}A_{3}(R)\,\left(R^{2}-(\sigma/2)^{2}\right)+O_{3}(h)\;,\label{eq:Qg03}\end{equation}
coincidentally $b_{2}=\pi\sigma^{2}$ is equal to the first non ideal
gas correction. Here $Q$ goes to zero as $h^{2}$, i.e., the system
lost two degrees of freedom. It is an expected property, which must
be valid independently of the dimension and even the number of particles,
$Q_{D}(R)\simeq h^{N}$ for $R$ large enough. After $h$ goes to
zero, $Q/h^{2}$ goes to zero as $\left(R-\sigma/2\right)$ it is
the caging limit. In the 2-HS-HWSP in $D=3$, three degrees of freedom
are lost when $R\rightarrow\sigma/2$ (see Eqs. (\ref{eq:q3}, \ref{eq:qDser2},
\ref{eq:QDser2})) as the system becomes caged, in the spherical pore
with an internal core they are lost in two steps.

The study of the systematic dependence on the dimension number of
the $h\rightarrow0$ limit is not easily available from Eq. (\ref{eq:Q2P_def}).
However, from Eq. (\ref{eq:Q2P_def01}) we find that the partition
function for $h\rightarrow0$ is $Q_{D}=h^{2}Q_{D/s}+O_{3}(h)$ with
$Q_{D/s}$ the configuration integral for both HS confined to the
surface of the $D$-dimensional sphere also known as the calottes
problem \cite{calotes}. By means of two stick transformation, the
$Q_{D/s}$ may be expressed as \begin{equation}
Q_{D/s}=S_{D}\left.\partial_{R1}\partial_{R2}Q_{D}(R_{1},R_{2})\right|_{R1=R2=R}=S_{D}\left.\partial_{R1}\partial_{R2}\int_{\sigma}^{R1+R2}W_{D}(r,R_{1},R_{2})r^{D-1}\, dr\right|_{R1=R2=R}\:,\label{eq:QDs_0}\end{equation}
where $Q_{D}(R_{1},R_{2})$ is the CI in Eq. (\ref{eq:QD_2}) but
with a different confinement radius for each particle, which is known
for $D=2,3$ \cite{Urru}. Here, we transform the $e_{\mathrm{A}}$,
$e_{\mathrm{B}}$ Heaviside functions on Eqs. (\ref{eq:QD}, \ref{eq:QD_1})
into Dirac delta functions through the derivatives $-\partial_{Ri}$,
sticking both HS on the surface. For practical purposes is much easier
to evaluate $Q_{D/s}$ from\begin{equation}
Q_{D/s}=A_{D}(R)\, S_{D-1}R^{D-1}\int_{0}^{\pi-d}\sin^{D-2}(\theta)\, d\theta\:,\label{eq:QDs_01}\end{equation}
\begin{equation}
Q_{D/s}=A_{D}^{2}(R)\, u_{D-2}(z)\:,\label{eq:QDs_02}\end{equation}
where $A_{D}(R)$ is the system volume and the minimum angular distance
between particles is $d=2\arcsin(z)$. Here $u_{D-2}(z)$ is the reduced
partition function which plays the same role that $q_{D}$ played
in the HWSP, then we define $q_{D/s}(z)\equiv u_{D-2}(z)$. As far
as the iterative construction of the $u_{D}(z)$ functions was previously
described we present the function for $D=1,\,2,\,3,\,4,$ and $5$
\begin{equation}
q_{1/s}(z)=1\:,\label{eq:q1s}\end{equation}
\begin{equation}
q_{2/s}(z)=\frac{2}{\pi}\arccos(z)\:,\label{eq:q2s}\end{equation}
\begin{equation}
q_{3/s}(z)=1-z^{2}\:,\label{eq:q3s}\end{equation}
\begin{equation}
q_{4/s}(z)=\frac{2}{\pi}\left[\arccos(z)+\sqrt{1-z^{2}}\, z(1-2z^{2})\right]\:,\label{eq:q4s}\end{equation}
\begin{equation}
q_{5/s}(z)=1-3z^{4}+2z^{6}\:,\label{eq:q5s}\end{equation}
an extra $1/2$ factor must be considered in Eq. (\ref{eq:q1s}) due
to a kind of ergodic to non-ergodic transition. The asymptotic behavior
may be accomplished with the approximate series representations for
$q_{D/s}$ \cite{Functions}\begin{equation}
q_{D/s}(z)=1-z^{D-1}C_{4}\left[1-\frac{(D-3)(D-1)}{2(D+1)}z^{2}+\frac{(D-5)(D-3)(D-1)}{2^{3}(D+3)}z^{4}+O_{6}(z^{2})\right]\:,\label{eq:qDsser1}\end{equation}
\begin{equation}
q_{D/s}(z)=(1-z^{2})^{(D-1)/2}C_{4}\left[1-\frac{(D-3)(D-1)}{2(D+1)}(1-z^{2})+O_{2}(1-z^{2})\right]\:,\label{eq:qDsser2}\end{equation}
where $C_{4}=\left[\frac{D-1}{2}\, B(\frac{D-1}{2},\frac{D-1}{2})\right]^{-1}$
and truncated terms in Eq. (\ref{eq:qDsser1}) are even powers in
$z$. A noticeable characteristic of both series is that they have
the same numerical coefficients and both are polynomial with $2(D-2)$
degree for odd dimension. The approximate series for $Q_{D/s}$ are
then\begin{equation}
Q_{D/s}=A_{D}^{2}(R)-A_{D}(R)\,2b_{D-1}+J\!\! J_{D}(R)\,2b_{D+1}\,\frac{D-3}{D-1}(2^{4}\pi)^{-1}-O_{D-5}(R)\:,\label{eq:QDsser1}\end{equation}
where we note that $b_{D-1}=2\pi a_{D-2}$ and the curvature coefficient
is $6\pi\delta_{D-2}^{(1)}$ in concordance with Eqs. (\ref{eq:aD},
\ref{eq:delta1D}), and\begin{equation}
Q_{D/s}=\sigma^{3(D-1)/2}C_{5}\,(R-\sigma/2)^{(D-1)/2}\left[1-\sigma^{-1}\frac{(D-1)(7D-9)}{2(D+1)}\,(R-\sigma/2)+\sigma^{-2}O(R-\sigma/2)^{2}\right]\:,\label{eq:QDsser2}\end{equation}
where $C_{5}=2^{2(D-1)}\mathtt{B}_{D}/D$ and the low density limit
implies a vanishing curvature limit too. From Eq. (\ref{eq:QDsser1})
we may obtain the first cluster integral correction due to the space
curvature for this non euclidean container or spherical boundary conditions.
We may compare the first cluster integral in the $(D-1)$-euclidean
space $2b_{D-1}$ with the cluster integral in the surface of a sphere
in $D$-dimensions $2b_{D/s}=(Q_{D/s}-A_{D}^{2}(R))/A_{D}(R)$, only
for $D=3$ we have $2b_{D/s}=2b_{D-1}$, for any dimension we obtain
$\underset{R\rightarrow\infty}{\lim}2b_{D/s}=2b_{D-1}$. Partition
function for $D=2,\,4,$ and $5$ are

\begin{equation}
Q_{2/s}=A_{2}^{2}(R)-A_{2}(R)\, S_{2}\frac{2R}{\pi}\arcsin(z)\:,\label{eq:Q2s}\end{equation}
\begin{equation}
Q_{4/s}=(\pi R^{3})^{2}(1-\arcsin(z)+z(1-2z^{2})(1-z^{2})^{1/2})\:,\label{eq:Q4s}\end{equation}
\begin{eqnarray}
Q_{5/s} & = & A_{5}^{2}(R)-A_{5}(R)\,\pi^{2}\sigma^{4}/2+A_{5}(R).R^{-2}\pi^{2}\sigma^{6}/12\nonumber \\
 & = & \left(\frac{4\pi}{3}\sqrt{2}\right)^{2}R^{2}(R-\frac{\sigma}{2})^{2}(R+\frac{\sigma}{2})^{2}(2R^{2}+\sigma^{2})\:.\label{eq:Q5s}\end{eqnarray}
Analyzing the way in which $Q_{D/s}$ goes to zero we obtain from
Eq. (\ref{eq:qDsser2}) that, \begin{equation}
LDF=(D-1)/2=(D+3)/2-2\:,\label{eq:ldf03}\end{equation}
i.e. two $LDF$ from $Q_{D}$ corresponds to the confinement of both
particles on the cavity surface. Now, we are able to extend the results
about $LDF$ in $Q_{D}$, $Q_{D/b}$ and $Q_{D/s}$ we may argue that
if we stick both particles between them and to the surface we find
\begin{equation}
LDF=(D-3)/2=(D+3)/2-3\:,\label{eq:ldf04}\end{equation}
where each stick-bond transformation has reduced the $LDF$ on one
unit. The idea is that the same final solid-like configuration can
be obtained in several ways, but its state properties can not depend
on the particular taken path. Therefore, to stick particles between
them, next stick each particle to the surface and finally cage the
system must produce the same CI (basically the system free energy)
that is obtained directly from the caging of the 2-HS-HWSP.

\begin{center}
\begin{table}[H]
\begin{centering}
\begin{tabular}{|c|c|ccccccc|}
\hline 
 &  &  &  &  & $D$ &  &  & \tabularnewline
\hline 
Coeff. & Eq. & -1 & 0 & 1 & 2 & 3 & 4 & 5\tabularnewline
\hline
\hline 
$S_{D}$ & (\ref{eq:Sangle}) & $-\pi^{-1}$ & $0$ & $2$ & $2\pi$ & $4\pi$ & $2\pi^{2}$ & $8\pi^{2}/3$\tabularnewline
\hline 
$2b_{D}/\sigma^{D}$ & (\ref{eq:2bgraph},\ref{eq:QDser1}) & $\pi^{-1}$ & $1$ & $2$ & $\pi$ & $4\pi/3$ & $\pi^{2}/2$ & $8\pi^{2}/15$\tabularnewline
\hline 
$2a_{D}/\sigma^{D+1}$ & (\ref{eq:QDser1}) & $-$ & $-$ & $1/2$ & $2/3$ & $\pi/4$ & $4\pi/15$ & $\pi^{2}/12$\tabularnewline
\hline 
$2\delta_{D}^{(1)}/\sigma^{D+3}$ & (\ref{eq:QDser1}) & $-$ & $-$ & $-$ & $1/60$ & $\pi/288$ & $\pi/378$ & $\pi^{2}/1536$\tabularnewline
\hline 
$C_{1,0}$ & $2^{D}$ & $1/2$ & $1$ & $2$ & $4$ & $8$ & $16$ & $32$\tabularnewline
\hline 
$C_{1,1}$ & (\ref{eq:qDser1}) & $-1/2$ & $0$ & $1$ & $2^{5}/(3\pi)$ & $9$ & $2^{10}/(15\pi)$ & $50$\tabularnewline
\hline 
$-C_{1,2}$ & (\ref{eq:qDser1}) & $0$ & $0$ & $0$ & $2^{4}/(15\pi)$ & $2$ & $2^{9}/(21\pi)$ & $25$\tabularnewline
\hline
$C_{2,0}$ & (\ref{eq:qDser2}) & $-$ & $0$ & $1/4$ & $2^{5}/(15\pi)$ & $3/2$ & $2^{10}/(105\pi)$ & $25/4$\tabularnewline
\hline
$C_{2,1}$ & (\ref{eq:qDser2}) & $-$ & $7/10$ & $1/2$ & $3/14$ & $-1/8$ & $-1/2$ & $-9/10$\tabularnewline
\hline
LDF & (\ref{eq:qDser2},\ref{eq:QDser2},\ref{eq:ldf}) & $1$ & $1.5$ & $2$ & $2.5$ & $3$ & $3.5$ & $4$\tabularnewline
\hline
$C_{3,0}$ & (\ref{eq:QDser2}) & $-$ & $0$ & $4$ & $2^{6}\pi/15$ & $8\pi^{2}/3$ & $2^{7}\pi^{3}/105$ & $4\pi^{4}/9$\tabularnewline
\hline
$C_{3,1}$ & (\ref{eq:QDser2}) & $-$ & $-17/10$ & $0$ & $19/14$ & $5/2$ & $7/2$ & $22/5$\tabularnewline
\hline
$C_{4}$ & (\ref{eq:qDsser1}) & $-$ & $-$ & $1/2$ & $2/\pi$ & $1$ & $16/(3\pi)$ & $3$\tabularnewline
\hline
$C_{5}$ & (\ref{eq:QDsser2}) & $-$ & $1$ & $1/2$ & $4/\pi$ & $4$ & $128/(3\pi)$ & $48$\tabularnewline
\hline
\end{tabular}
\par\end{centering}

\caption{Several constant coefficients of the series representation of CIs
and other functions. Some values for $D=-1$ and $0$ correspond to
a suitable limiting behavior.\label{tab:1}}

\end{table}

\par\end{center}

\section{Mechanical equilibrium and Pressure tensor\label{sec:Pressure-ten}}

To make a more complete and microscopic characterization of the system
it is necessary to evaluate the pressure tensor. Briefly, pressure
tensor, density and external potential are related as a consequence
of the mechanical equilibrium by\begin{equation}
\nabla\cdot\mathbf{P}+\rho\nabla\psi=0\:,\label{eq:Leqdiff00}\end{equation}
where the explicit position dependence of the magnitudes has been
dropped. In any inhomogeneous system the tensor can be split into\begin{equation}
\mathbf{P}=\beta^{-1}\rho\,\mathbf{Id}+\mathbf{P}^{U}\;,\label{eq:Pt01}\end{equation}
where $\mathbf{Id}$ is the $D\times D$ identity matrix and $\mathbf{P}^{U}$
is the interaction part of the tensor \cite{Mor}. In systems with
spherical symmetry $\mathbf{P}^{U}$ is diagonal with only two different
components \cite{Blok1,Poni} the normal and tangential components
$P_{N}^{U}$ and $P_{T}^{U}$\begin{equation}
\mathbf{P}^{U}=P_{N}^{U}\hat{\mathbf{r}}\hat{\mathbf{r}}+P_{T}^{U}(\hat{\boldsymbol{\varphi}}\hat{\boldsymbol{\varphi}}+\hat{\boldsymbol{\theta}}_{1}\hat{\boldsymbol{\theta}}_{1}+...+\hat{\boldsymbol{\theta}}_{D-2}\hat{\boldsymbol{\theta}}_{D-2})\;,\label{eq:Pt02}\end{equation}
where $\hat{\boldsymbol{\varphi}},\hat{\boldsymbol{\theta}}_{1},...,\hat{\boldsymbol{\theta}}_{D-2}$
are the angular versors. Then, for systems with spherical symmetry
Eq. (\ref{eq:Leqdiff00}) may be written as\begin{equation}
\partial_{r}P_{N}+\frac{D-1}{r}(P_{N}-P_{T})+\rho\partial_{r}\psi=0\:.\label{eq:Leqdiff01}\end{equation}
Further simplifications apply to a hard wall container, there $\rho(r)$
is discontinuous at $r=R$ but $\varrho(r)$ is not (see Eq. (\ref{eq:rhovar})),
we obtain from Eq. (\ref{eq:Leqdiff01})\begin{equation}
\partial_{r}P_{N}+\frac{D-1}{r}(P_{N}-P_{T})=-\beta^{-1}\delta(r-R)\,\varrho\:,\label{eq:Leqdiff02}\end{equation}
\begin{equation}
\partial_{r}P_{N}^{U}+\frac{D-1}{r}(P_{N}^{U}-P_{T}^{U})=-\beta^{-1}e_{\mathrm{A}}(r)\,\partial_{r}\varrho\:,\label{eq:Leqdiff03}\end{equation}
both equations are equivalent to Eq. (\ref{eq:Leqdiff01}) and then
necessary conditions for an acceptable pressure tensor definition.
Nevertheless they are formally strongly different. The inhomogeneous
term in the differential equation (\ref{eq:Leqdiff02}) shows at the
boundary a divergent singular contribution to the total pressure components
with zero contribution from positions inside and outside the boundary.
Although, inhomogeneous term in Eq. (\ref{eq:Leqdiff03}) shows at
the boundary a discontinuous singular contribution to $P^{U}$ components,
with non zero contribution from inside points. We will regress to
this issue in a forthcoming paragraph.

\begin{figure}[h]
\begin{centering}
\includegraphics[clip,width=7cm]{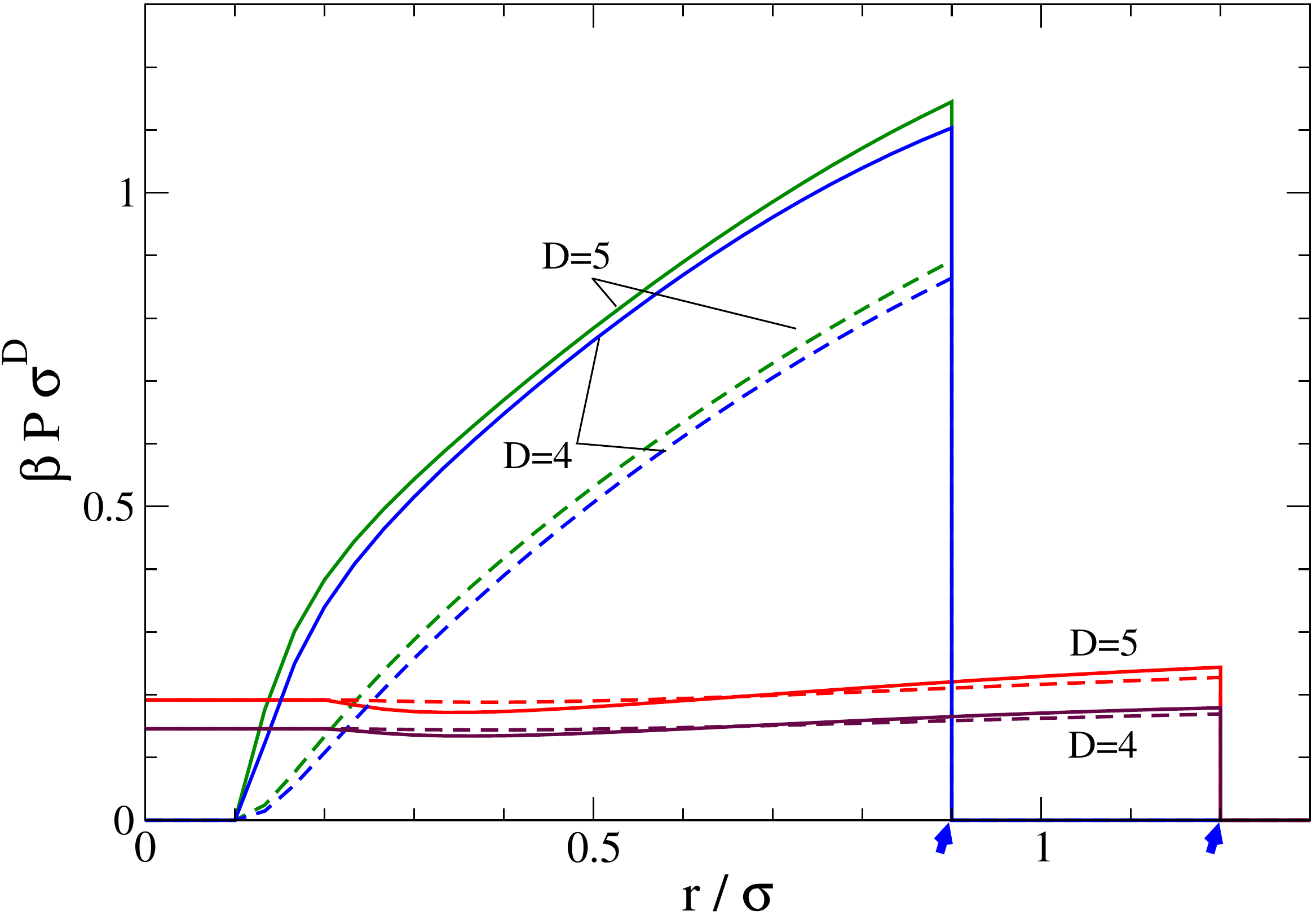}
\par\end{centering}

\caption{(color online) Pressure tensor components. Continuous line corresponds
to $P_{N}$ and dashed line to $P_{T}$. Two cavity sizes $R/\sigma=0.9$
and $R/\sigma=1.2$, and two different dimension $D$ are shown. The
arrows indicate $R/\sigma$ values.\label{fig:Pressureposition}}

\end{figure}

\begin{figure}[h]
\centering{}\includegraphics[clip,width=7cm]{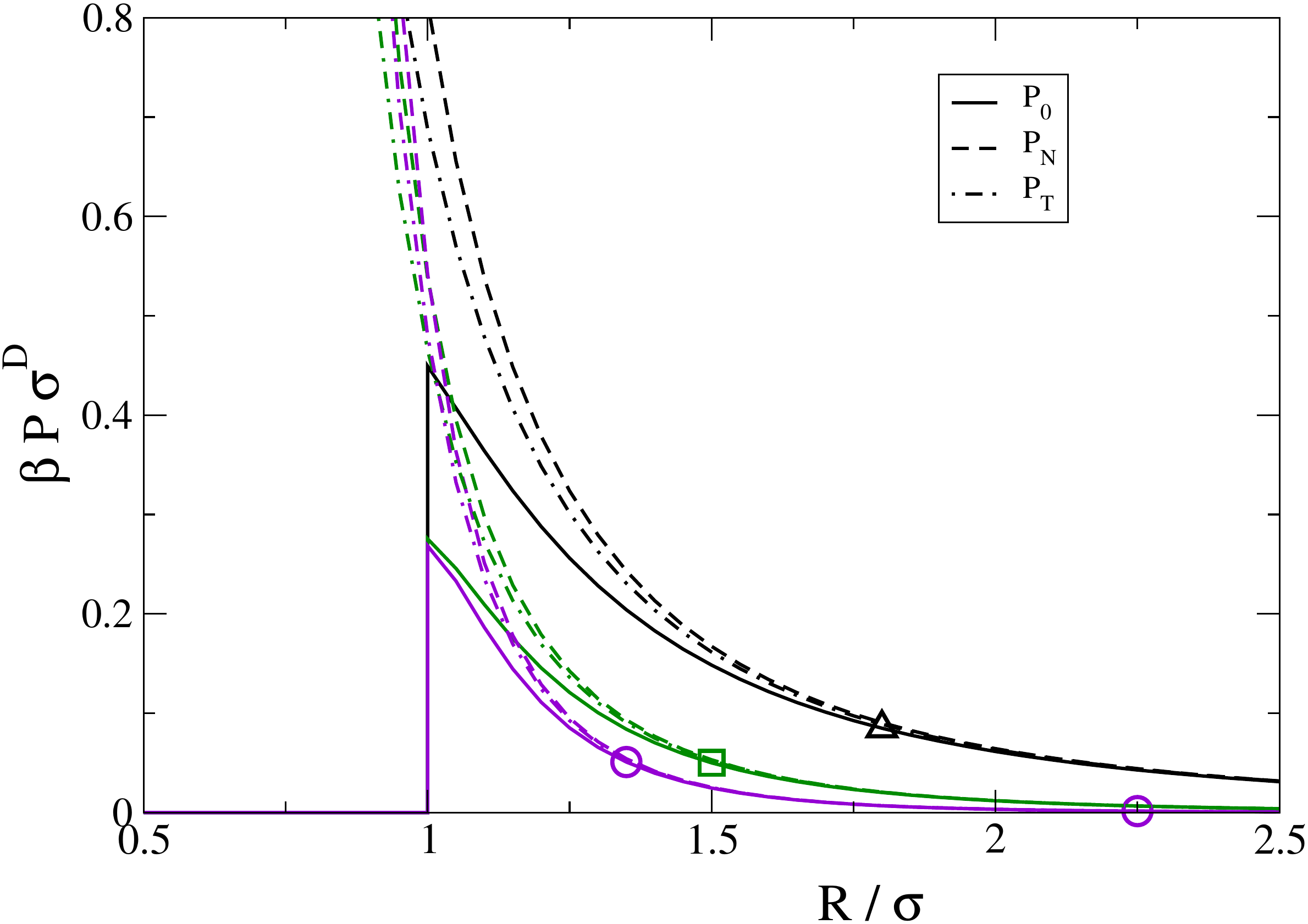}\caption{(color online) Characteristic values of pressure tensor. Pressure,
at the central plateau $P_{0}$ (continuous line) and contact value
($r=R$) of both components $P_{N}$ and $P_{T}$ (dashed line and
dash-dotted line) as a function of pore size. From top to bottom dimensions
$D=3,\,5,$ and $7$ in triangles, squares and circles.\label{fig:pressure-hom-cont}}

\end{figure}

Turning now to the interaction part of pressure $P^{U}$, is known
that its detailed expression is non unique. Particularly, different
possible definitions of $P^{U}$ produce different values of pressure
tensor in inhomogeneous fluids \cite{Blok1}. In this work we adopt
a pressure tensor definition extensively utilized in MD simulations
\cite{Mor}, the components of the pressure tensor for the two body
system are\begin{equation}
P_{ab}^{U}(r)\equiv\left\langle r_{\mathrm{AB}}^{a}F_{\mathrm{AB}}^{b}(\mathbf{r}_{\mathrm{AB}})\delta(\mathbf{r}-\mathbf{r}_{\mathrm{A}})\right\rangle \:,\label{eq:Pmd}\end{equation}
been $r^{a}=\mathbf{r}\cdot\hat{\mathbf{a}}$ and $F_{\mathrm{AB}}^{b}(\mathbf{r})=-\frac{d\phi}{dr}\frac{r^{b}}{r}$
the component of the force between particles in the $\hat{\mathbf{b}}$
direction (see Eq. (3.20) in Ref. \cite{Mor}). It has been argued
that Eq. (\ref{eq:Pmd}) can be obtained from the Irving Kirkwood
pressure tensor with the assumption of short range interaction \cite{Mor}.
Interestingly, the adopted definition for the pressure is an intrinsic
two bodies emergent property which depends only on the position of
two particle coordinates and does not depend on any choice of an integration
path. In the present problem is important to notice that position
of both bodies are the point of pressure evaluation $\mathbf{r}=\mathbf{r}_{\mathrm{A}}$
and the integrated position $\mathbf{r}_{\mathrm{B}}$ in the mean
value of Eq. (\ref{eq:Pmd}). It is a desirable property of $\mathbf{P}^{U}(\mathbf{r})$
in any few body system that the microscopic configurations with zero
probability to find a particle in position $\mathbf{r}$ must not
contribute to the pressure. This property is not accomplished by the
Irving-Kirkwood choice for the pressure tensor. Equation (\ref{eq:Pmd})
may be written as \begin{equation}
P_{ab}^{U}(r)=(\beta Q_{D})^{-1}e_{\mathrm{A}}(r)\intop e'_{\mathrm{AB}}e_{\mathrm{B}}\frac{r_{\mathrm{AB}}^{a}r_{\mathrm{AB}}^{b}}{|r_{\mathrm{AB}}|}d\mathbf{r}_{\mathrm{B}}\:,\label{eq:Pmd1}\end{equation}
where $\mathbf{r}_{\mathrm{AB}}=\mathbf{r}-\mathbf{r}_{\mathrm{B}}$
and $e'_{\mathrm{AB}}=\delta(r_{\mathrm{AB}}-\sigma)$. Changing the
integration variable to $\mathbf{r}_{\mathrm{AB}}$ and integrating
over the radial coordinate one gets\begin{equation}
P_{ab}^{U}(r)=(\beta Q_{D})^{-1}\sigma^{D}e_{\mathrm{A}}(r)\intop e_{\mathrm{B}}(r_{\mathrm{B}})\hat{r}_{\mathrm{AB}}^{a}\hat{r}_{\mathrm{AB}}^{b}d\Omega_{\mathrm{AB}}\:,\label{eq:Pmd2}\end{equation}
with $r_{\mathrm{B}}=\left|\mathbf{r}-\sigma\,\hat{\mathbf{r}}_{\mathrm{AB}}\right|$
and $\hat{r}^{b}=\hat{\mathbf{r}}\cdot\hat{\mathbf{b}}$. The two
independent components of this tensor are\begin{equation}
P_{N}^{U}(r)=(\beta Q_{D})^{-1}e_{\mathrm{A}}(r)\,\sigma^{D}S_{D-1}\,\intop e_{\mathrm{B}}\cos^{2}\theta\sin^{D-2}\theta\, d\theta\:,\label{eq:PmdN}\end{equation}
\begin{equation}
P_{T}^{U}(r)=(\beta Q_{D})^{-1}e_{\mathrm{A}}(r)\,\sigma^{D}S_{D-1}(D-1)^{-1}\,\intop e_{\mathrm{B}}\sin^{D}\theta\, d\theta\:,\label{eq:PmdT}\end{equation}
where the integration interval $(0,\pi)$ may be reduced to $(\theta_{min},\pi)$
by the effect of $e_{\mathrm{B}}$, with $\theta_{min}=0$ if $r<R-\sigma$,
$\theta_{min}=\pi$ if $|r-\sigma|>R$, and $\theta_{min}=\pi-\arccos(-R^{2}+\sigma^{2}+r^{2})/(2r\sigma)$
otherwise. Both integrals expressed in terms of known functions are\begin{equation}
P_{N}^{U}(r)=(\beta Q_{D})^{-1}e_{\mathrm{A}}(r)\,2^{-1}V_{D}(\sigma)\:\left[D\, w_{D-2}(\cos\alpha)-(D-1)\, w_{D}(\cos\alpha)\right]\:,\label{eq:PmdN3}\end{equation}
\begin{equation}
P_{T}^{U}(r)=(\beta Q_{D})^{-1}e_{\mathrm{A}}(r)\,2^{-1}V_{D}(\sigma)\: w_{D}(\cos\alpha)\:,\label{eq:PmdT3}\end{equation}
where $\cos\alpha=1$ if $r>R+\sigma$, $\cos\alpha=-sign(R-\sigma)$
if $r<R-\sigma$ and $\cos\alpha=(-R^{2}+\sigma^{2}+r^{2})/(2r\sigma)$
if $R-\sigma\leq r\leq R+\sigma$. Figure \ref{fig:Pressureposition}
shows both pressure tensor components as a function of position into
the cavity. We observe that when the pore is big enough, $R>\sigma$,
a pressure plateau develops at the center of the pore in the range
$0\leq r<R-\sigma$. In the region $R-\sigma\leq r<R$ the inhomogeneous
pressure region develops. The shape of pressure tensor and density
profiles are simply correlated (see Fig. 2 in \cite{Urru}), their
constant value plateau and inhomogeneous regions coincide, even more,
when plateau density becomes null, pressure goes to zero too. %
In the homogeneous region the normal and tangential components of
the pressure tensor become equal, and the pressure tensor reduces
to a constant scalar pressure $P_{0}$ \cite{Blok1}. In this case
Eqs. (\ref{eq:PmdN3}, \ref{eq:PmdT3}) lead to\begin{equation}
P_{N,T}^{U}(0)=P_{0}^{U}=(\beta Q_{D})^{-1}2b_{D}\:,\label{eq:PmdNTu}\end{equation}
finally, according to Eq. (\ref{eq:Pt01}) the pressure in the homogeneous
density plateau is\begin{equation}
\beta P_{0}=\rho_{0}+\frac{2b_{D}}{Q_{D}}=2(V_{D}-b_{D})/Q_{D}\:,\label{eq:P0}\end{equation}
for $R\geq\sigma$ while $\beta P_{0}=0$ for $\sigma/2\leq R<\sigma$.
The value of the pressure tensor at contact can be obtained from Eqs.
(\ref{eq:PmdN3}, \ref{eq:PmdT3}) with $r=R$ and $\cos\alpha=z=\sigma/(2R)$
but no further simplification may be done. In Fig. \ref{fig:pressure-hom-cont}
we plotted the characteristic values of the pressure tensor. There
we can observe that the maximum of $P_{0}$ is attained at smaller
$R$ than the maximum of the density (see Fig. \ref{fig:density-min-max}).
We have verified that in the inhomogeneous region $abs(R-\sigma)\leq r\leq R$,
Eq. (\ref{eq:Leqdiff03}) with $P_{N,T}$ from Eqs. (\ref{eq:PmdN3},
\ref{eq:PmdT3}) is false. Two probable reasons may be argued, the
invalid short range hypothesis for the HS potential for nonuniform
density, and/or the incorrect pressure tensor definition of Eq. (\ref{eq:Pmd})
due to the hard wall boundary condition (see Eq.(3.5) in \cite{Poni}).
As far as the mechanical equilibrium of Eq. (\ref{eq:Leqdiff03})
is still valid in the homogeneous plateau, and pressure should not
strongly depend on the definition details in this region, we accept
the validity of the obtained pressure tensor for $0<r<abs(R-\sigma)$.
It is particularly interesting to analyse the point $r=R$ where we
find $\partial_{r}\varrho>0$. We may note that the discontinuous
behaviour of $P_{N}^{U}$ (but not of $P_{T}^{U}$) at $r=R$ violates
Eq. (\ref{eq:Leqdiff03}). The existence of the Heaviside factor $e_{A}(r)$
on the right hand side of Eq. (\ref{eq:Leqdiff03}) must be balanced
with the same global factor at the left side, and forbids the appearance
of an uncompensated Dirac delta. However, the derivative of a discontinuous
$P_{N}^{U}$ just produce a singular Dirac delta at $r=R$. If we
assume the validity of this equation one obtains that the discontinuity
is completely unphysical. Thus, we need the continuity of the interaction
normal component of the pressure tensor to overcome the mismatch.

\section{The Equation of state and the Laplace Equation\label{sec:EOS}}

\begin{figure}
\begin{centering}
\includegraphics[width=8cm]{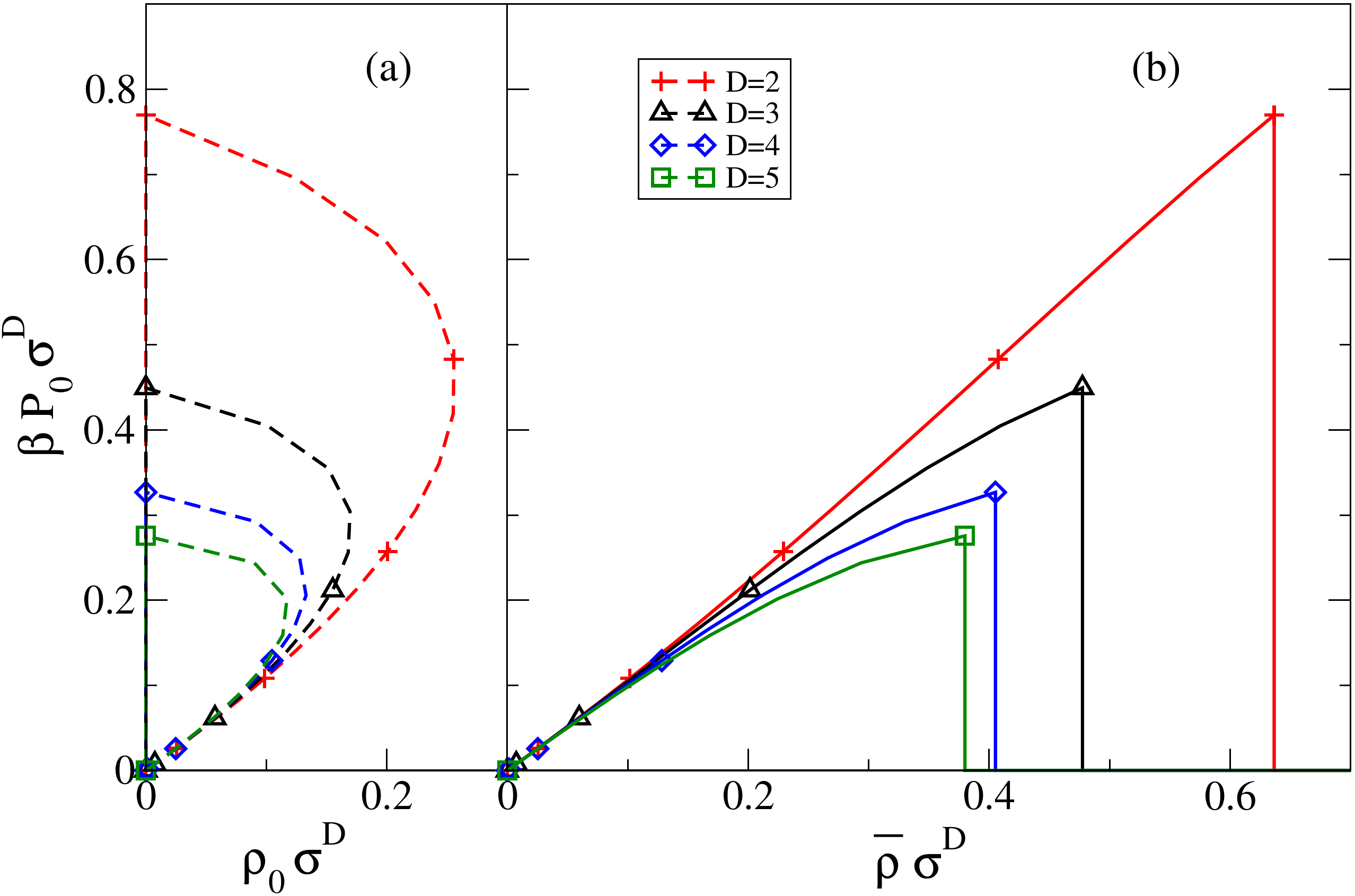}
\par\end{centering}

\caption{(color online) Pressure in the homogeneous plateau. From top to bottom
$D=2,3,4,$ and $5$. Two different choices of density parameter are
depicted. At left hand side (a), in dashed line we adopt the $\rho_{0}$
as density parameter. At right hand side (b), in continuous line we
adopt the rough density $\bar{\rho}$.\label{fig:EOS}}

\end{figure}

We are interested in the equation of state (EOS) of the 2-HS-HWSP,
i.e. the thermodynamic description of the complete inhomogeneous system.
We adopt a point of view usually taken in spherical droplets which
makes a description in terms of the properties in the homogeneous
regions of the system. In this section we drop any explicit unnecessary
$D$ subindex and dependence on cavity size $R$, then $V=V_{D}(R)$,
$Q=Q_{D}(R)$ and so on. The properties in the homogeneous region
may be found from Eqs. (\ref{eq:rho0}, \ref{eq:P0}). From these
Eqs., for $R>\sigma$ we obtain \begin{eqnarray}
\frac{\beta P_{0}}{\rho_{0}} & = & 1+\frac{b}{2}\,\frac{\bar{\rho}}{1-b\bar{\rho}}\:,\label{eq:P0/rho0}\\
 & = & 1+\frac{1}{2}\,(V/2b-1)^{-1}\:,\label{eq:P0/rho0a}\end{eqnarray}
where the compressibility factor at the constant density plateau is
a simple function of $\bar{\rho}$ or $V$. We may note that both
expressions diverge at $R=\sigma$ and $V=2b$. For this pore size
the plateau of constant density vanishes and $\rho_{0}=0$. Interestingly,
in Eq. (\ref{eq:P0/rho0a}) we have found the bulk van der Waals EOS
without the attractive term, the only flavor remaining the two body
system is a $1/2$ factor. Owing to Eq. (\ref{eq:P0/rho0}) does not
depend on geometrical parameters which resembles the cavity's spherical
symmetry it should be valid also for 2-HS confined in pores with other
geometries. In Eqs. (\ref{eq:P0/rho0}, \ref{eq:P0/rho0a}) $P_{0},\:\rho_{0}$
and $\bar{\rho}$ (or $V$) are not independent variables, even, they
are functions of system's size $R$. The Eq. (\ref{eq:P0/rho0}) can
be written in two forms which resemble the contact theorem for the
bulk homogeneous HS system (see Eq. (2.5.26) in \cite{Hansen} and
also \cite{Reiss})\begin{eqnarray}
\beta P_{0} & = & \rho_{0}+b\bar{\rho}^{2}\,\frac{g(\sigma)}{w(z)}\:,\label{eq:P0/rhobisa}\\
 & = & \rho_{0}+\frac{b}{2}\rho_{0}^{2}\,\frac{q(z)}{(1-2z)^{2}}\:,\label{eq:P0/rhobis}\end{eqnarray}
where in fact $g(\sigma)/w(z)=(2q(z))^{-1}$ and then $g(\sigma)/w(z)\rightarrow1/2$
when $R\rightarrow\infty$. The Eqs. (\ref{eq:P0/rhobisa}, \ref{eq:P0/rhobis})
make sense only for $0\leq z<1/2$ or $\sigma<R\leq\infty$, where
for the smallest pore ($R=\sigma$ and $z=1/2$) $\rho_{0}=0$. In
such a case, the term which multiply $\bar{\rho}^{2}$ in Eq. (\ref{eq:P0/rhobisa})
takes a finite value, but, the term which multiply $\rho_{0}^{2}$
in Eq. (\ref{eq:P0/rhobis}) diverges. However, Eq. (\ref{eq:P0/rhobis})
goes to zero. This shows an only apparent different behaviour, because
Eqs. (\ref{eq:P0/rhobisa}, \ref{eq:P0/rhobis}) are different representations
of the same equation. At large $R$ the overall and plateau densities
are related by\begin{equation}
\rho_{0}\simeq\bar{\rho}-\frac{a}{2}\,\bar{\rho}^{3}A(R)\simeq\bar{\rho}-C_{6}\,\bar{\rho}^{2+1/D}\:,\label{eq:rho0Ser}\end{equation}
with $C_{6}=(D^{D-1}2^{-1}S_{D})^{1/D}a_{D}$ and $A(R)\sim\rho_{0}^{-(D-1)/D}\sim\bar{\rho}^{^{-}(D-1)/D}$
for $D>1$. Polynomial expressions for the pressure as a function
of density, may be found by considering the firsts terms of $P_{0}$
as a density power series\begin{eqnarray}
\beta P_{0} & \simeq & \rho_{0}+\frac{b}{2}\rho_{0}^{2}+\frac{b^{2}}{2}\rho_{0}^{3}+\frac{b}{4}\,\rho_{0}^{4}\left(2b^{2}+a\, A(R)\right)\:,\label{eq:P0eos1-3}\\
 & \simeq & \bar{\rho}+\frac{b}{2}\,\bar{\rho}^{2}+\frac{1}{2}\bar{\rho}^{3}\left(b^{2}-a\, A(R)\right)\:.\label{eq:P0eos2}\end{eqnarray}
Here we choose two density parametrization, the plateau density and
rough density, $\rho_{0}$ and $\bar{\rho}$ respectively. The expansions
in Eqs. (\ref{eq:P0eos1-3}, \ref{eq:P0eos2}) end at the order of
the first signature of the inhomogeneity. Both equations show that
the first correction to the ideal gas behaviour is positive and involves
the expected closed system correction \cite{Hub}. Besides, the Eqs.
(\ref{eq:P0eos1-3}, \ref{eq:P0eos2}) involve terms with higher powers
in the density than two. This feature may sound conflicting in a two
body system, but in fact, it is a direct consequence of the fixed
number of particles that characterizes the Canonical Ensemble approach.
In Fig. \ref{fig:EOS} we plot the pressure in the homogeneous plateau
as a function of both density parameters. In Fig \ref{fig:EOS} (a)
we adopt the plateau density $\rho_{0}$. There, an important characteristic
is apparent, for small cavities $\sigma/2<R<\sigma$ we obtain $\rho_{0}=0$
and $P_{0}=0$ (see also Figs. \ref{fig:density-min-max} and \ref{fig:Pressureposition}).
The anomalous bi-valued behaviour of $P_{0}(\rho_{0})$ is a consequence
of the non monotonic behavior of the density in the homogeneous region
as was shown in Fig. \ref{fig:density-min-max}. In Fig. \ref{fig:EOS}
(b) we adopt the rough density $\bar{\rho}$. For $P_{0}(\bar{\rho})$
we find a simpler general dependence with a monotonic behaviour until
$\bar{\rho}=2/V(1)$ is reached. Note that the maximum attainable
density is $\bar{\rho}_{max}=2/V(1/2)$. In both Figs. \ref{fig:EOS}
(a) and (b) the pressure attains its maximum when the density plateau
disappears at $R=\sigma$ and then pressure drops discontinuously
to zero. %
Interestingly, both used density parameters $\{\rho_{0},\bar{\rho}\}$
are usually utilized to describe the behaviour of macroscopic fluid
systems. From an opposite point of view, we may concentrate in the
external force and on contact properties. From the wall theorem \cite{Blok3,Lovet}
the total scalar force between the wall and the HS system in a pressure
form is\begin{equation}
d_{R}\ln(\tilde{Q})\cdot A(R)^{-1}=\beta P_{W}=\rho_{c}\:,\label{eq:PW}\end{equation}
with $\rho_{c}$ from Eq (\ref{eq:rhoc}). The general features of
$\beta P_{W}(R)$ for $D=2$ and $3$ can be seen in Fig. 1 of Ref.
\cite{Urru}. There, the basic systematic behaviour with increasing
dimensionality is apparent. As a consequence of the chain rule of
the derivative and the sticky bond transformation the numerator of
$\rho_{c}$ is the CI of one HS in a diminished volume. The allowed
volume for the HS particle is the free volume available when the other
HS is sticked to the surface. The asymptotic behaviour of $P_{W}$
may be obtained from Eqs. (\ref{eq:QDser1}, \ref{eq:QDser2}). For
$D=2$ and $3$ it was studied in \cite{Urru}. In the caging limit,
when $R\rightarrow\sigma/2$, $P_{W}$ diverges as\begin{equation}
\beta P_{W}\simeq\frac{D+3}{2\, A(\sigma/2)}\,(R-\frac{\sigma}{2})^{-1}\:,\label{eq:PWasym}\end{equation}
\begin{equation}
\frac{\beta P_{W}}{\bar{\rho}}\simeq\frac{D+3}{4D}\,(V/V_{0}-1)^{-1}\:,\label{eq:PWasym2}\end{equation}
where $V_{0}=V(\sigma/2)$ and the expressions in Eqs. (\ref{eq:PWasym},
\ref{eq:PWasym2}) are consistent to order minus one in $(R-\frac{\sigma}{2})$
and $(V/V_{0}-1)$. The same power dependence of the compressibility
factor was found for the caging limit of N-HS systems under periodic
boundary conditions \cite{Sals}. A comparison between Eqs. (\ref{eq:P0/rho0a})
and (\ref{eq:PWasym2}) shows an interesting similarity at $D=3$.

\begin{figure}[b]
\centering{}\includegraphics[clip,width=7cm]{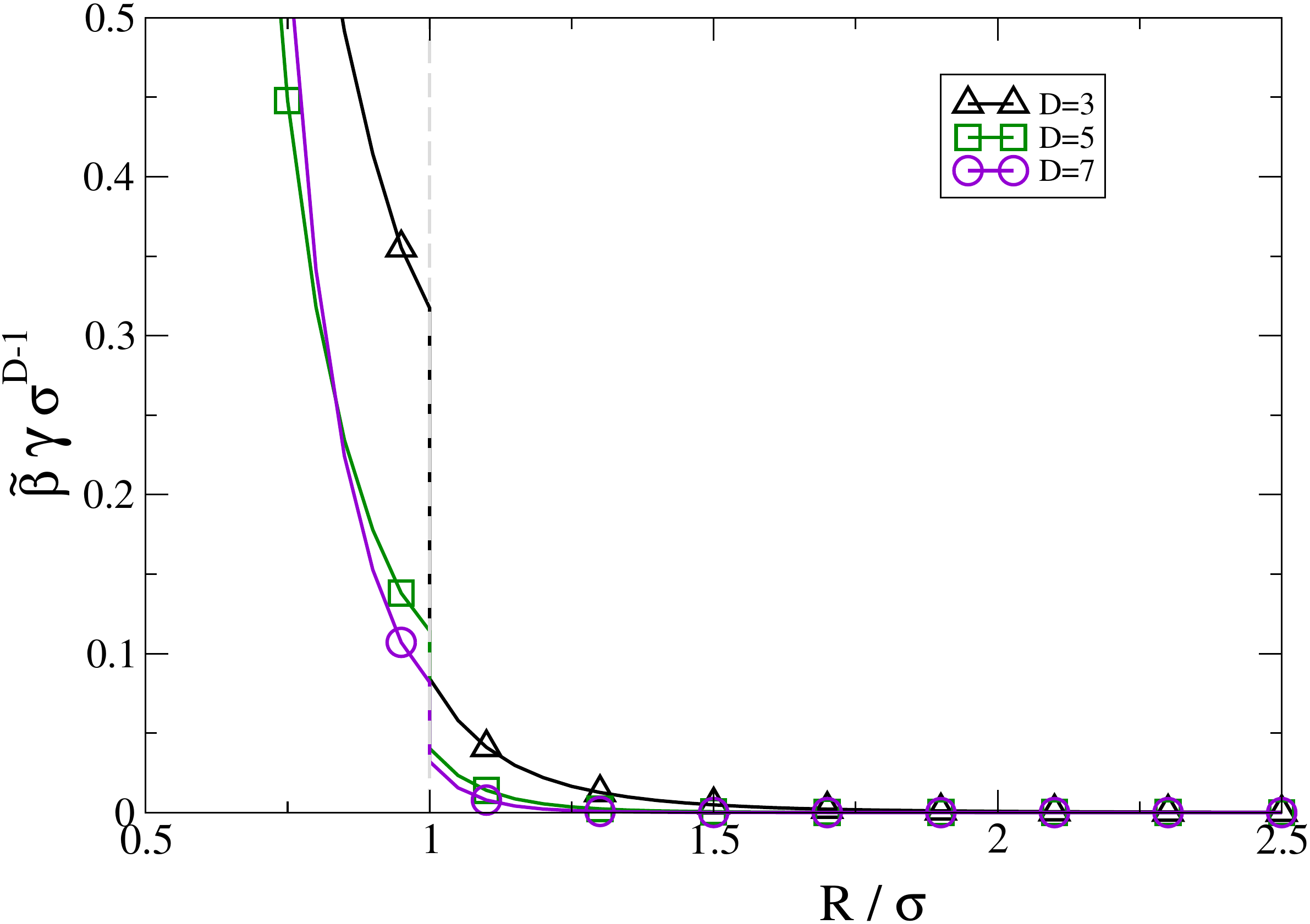}\caption{(color online) Function $\tilde{\gamma}$, its dependence on pore
radius for several dimensions. The dashed line shows the discontinuity
at $R/\sigma=1$. \label{fig:Tension}}

\end{figure}

Finally, we shall study the surface tension of the system. Due to
the failure of the obtained pressure tensor we can not evaluate a
microscopic expression for the surface tension, however we may adopt
a macroscopic approach. Identifying the radius of the dividing interface
with $R$ we define $\tilde{\gamma}$ by \begin{equation}
\beta\mathrm{j}\tilde{\gamma}(R)=\rho_{c}-\beta P_{0}=\beta\Delta P\:,\label{eq:SurfTen01}\end{equation}
where at the left hand side appears the pressure difference $\Delta P=P_{W}-P_{0}$.
If we identify $\tilde{\gamma}$ with the surface tension, the Eq.
(\ref{eq:SurfTen01}) is the original version of Laplace equation
applied to a spherical substrate-fluid interface \cite{Blok3}. Besides,
for $D=1$ or $R\rightarrow\infty$ we obtain the planar equilibrium
condition $\rho_{c}=\beta P_{0}$. Figure \ref{fig:Tension} shows
$\tilde{\gamma}$ for different radii and several dimensions $D>1$.
Some general features are: it is a positive defined quantity; at large
$R$ the value of $\tilde{\gamma}$ goes to zero with an expected
power law dependent on $D$; in the opposite, as $R/\sigma\rightarrow1/2$
$\tilde{\gamma}$ increases indefinitely as $\rho_{c}$ do. It can
be observed a finite jump at $R/\sigma=1$ due to a discontinuity
in $P_{0}$. Probably such discontinuities in $P_{0}$ and $\tilde{\gamma}$
are unphysical artifacts of the adopted definition for $P_{ab}^{U}(r)$.%
 For $R>\sigma$ we have\begin{eqnarray}
\beta\mathrm{j}\tilde{\gamma} & = & 2\left[V-(1-c(z))\, b-V+b\right]/Q\:,\nonumber \\
 & = & 2b\, c(z)/Q\:,\label{eq:SurfTen02}\end{eqnarray}
\begin{equation}
c(z)=1-w_{D}(z)-(2z)^{-D}w_{D}(1-2z^{2})\:,\label{eq:SurfC01}\end{equation}
where $0<c(z)<1$ and the most relevant terms of $c(z)$ are \begin{equation}
c(z)\simeq(D-1)\,\mathtt{B}_{D}z\,\left[2(D+1)^{-1}-\frac{D-3}{3(D+3)}z^{2}\right]\:,\label{eq:SurfC01ser}\end{equation}
where for $D=3$ we obtain the somewhat surprising result $c(z)=z\,3/4$.
To first non null order in density and curvature we obtain \begin{equation}
\beta\tilde{\gamma}\simeq\frac{a}{2}\bar{\rho}^{2}-\mathsf{\delta}_{D}^{(1)}2^{-1}(D-1)(D-3)\, R^{-2}\bar{\rho}^{2}\:,\label{eq:SurfTser01}\end{equation}
where first term does not show any curvature dependence. In Addition,
to the same order of accuracy in density we may replace $\bar{\rho}\rightarrow\rho_{0}$.
We identify the first density-non-curvature term as $\beta\gamma_{flat}=a/2\,\bar{\rho}^{2}$.
To first non null order we obtain \begin{equation}
\tilde{\gamma}/\gamma_{flat}-1\simeq-\left(\frac{\sigma}{R}\right)^{2}\,\frac{(D-3)(D+1)}{24\,(D+3)}\:,\label{eq:SurfTser02}\end{equation}
where the right hand term becomes null for $D=3$. Unfortunately $\tilde{\gamma}$
is not the surface tension but in fact it is an excess free work.
A better proposal for the definition of the surface tension $\gamma$
is a refined version of the Laplace equation \cite{Hend2} \begin{equation}
\partial_{R}\gamma+\mathrm{j}\gamma=\Delta P\:,\label{eq:SurfTen03}\end{equation}
where $\Delta P=\mathrm{j}\tilde{\gamma}$. Definition from Eq. (\ref{eq:SurfTen03})
separates an explicit curvature dependent term from $\Delta P$ \cite{Blok3}.
The asymptotic behaviour of $\Delta P$ is essentially described in
Eq. (\ref{eq:SurfTser01}), from that we obtain for the surface tension
\begin{equation}
\gamma/\gamma_{flat}-1\simeq-\left(\frac{\sigma}{R}\right)^{2}\,\frac{(D-1)(D+1)}{24\,(D+3)}\:,\label{eq:SurfTser03}\end{equation}
where now, the right hand term becomes null at $D=1$. We may study
the same system from a different radii which is an interesting point
of view principally for non sharp interfaces usually found in fluid
droplets. Being $R''=R+\varepsilon$ and following Henderson \cite{Hend2}
we define, \begin{equation}
\mathrm{j}(R)\,\beta\gamma(R)=\mathrm{j}(R'')\,\beta\gamma(R,R'')+\beta\partial_{R''}\gamma(R,R'')\:,\label{eq:SurfTeps01}\end{equation}
its solution is\begin{equation}
\beta\gamma(R,R'')=\beta\gamma(R)\,\frac{1}{D}\left\{ (D-1)\,\left(\frac{R}{R''}\right)^{-1}+\left(\frac{R}{R''}\right)^{D-1}\right\} \:,\label{eq:SurfTesp02}\end{equation}
and then\begin{equation}
\gamma(R,R'')/\gamma(R)-1=\frac{D-1}{2}\,\left(\frac{\varepsilon}{R}\right)^{2}+O_{3}(\frac{\varepsilon}{R})\:,\label{eq:SurfTesp03}\end{equation}
showing that first correction is second order in $\varepsilon/R$
and positive for $D>1$.

We expect that, several of the above expressions may be generalized
for any number of particles $N$ with the addition of the overall
$N$-dependent correction factor \cite{Hub}. Here we do not demonstrate
but merely suggest that, the Eq. (\ref{eq:SurfTser01}) should be
multiplied by $2(1-N^{-1})$ at right hand side \cite{Kratky} to
become valid to the same order in density and curvature. The same
modification applies to $\gamma_{flat}$. However, this prefactor
does not modify the ratios in Eqs. (\ref{eq:SurfTser02}, \ref{eq:SurfTser03}).
Therefore, for a large enough $R$ we obtain for the $N$-HS-HWSP
system in $D$ dimensions\begin{equation}
\beta\gamma_{flat}\simeq a\,(1-N^{-1})\bar{\rho}^{2}\:,\label{eq:SurfTflatN}\end{equation}
\begin{equation}
\gamma/\gamma_{flat}-1\simeq-\left(\frac{\sigma}{R}\right)^{2}\,\frac{(D-1)(D+1)}{24\,(D+3)}\:,\label{eq:SurfTserN}\end{equation}
both are new results even in the most interesting case of $D=3$.
Furthermore, the Eq. (\ref{eq:SurfTflatN}) represents the first term
of the density power series of the surface tension for any hard wall
cavity. In addition, Eq. (\ref{eq:SurfTserN}) applies too for the
\textit{conjugate system} of a hard wall spherical core surrounded
by a HS fluid. Finally, we are interested in establish some relations
between the studied system and the open (grand canonical ensemble)
system of HS in contact with a hard spherical wall. The low density
limit of an inhomogeneous open system was studied by Bellemans and
Sokolowski-Stecki \cite{Bell,Soko} who found the surface tension
virial series and apply it to the HS system in contact with a planar
hard wall (cite I Ref. \cite{Bell} and I, III Ref. \cite{Soko})
and contained in a HWSP (cite III in Ref. \cite{Bell}) in $D=3$.
In the HS inhomogeneous fluid in a HWSP the $\gamma$ first power
density coefficient and zero order in curvature is the same that appears
when a planar wall is studied (from Area term in Eq.(\ref{eq:QDser1}))
$\beta_{\Omega2}^{Bell}=W_{2}^{S-S}/A_{D}(R)=a_{D}=b_{D+1}(2\pi)^{-1}$
which may be found by taking $N\rightarrow\infty$ at Eq. (\ref{eq:SurfTflatN}).
The first curvature correction of this should be $\mathsf{\delta}_{D}^{(0)}$.
Bellemans obtained that in three dimensions this constant is null,
$\mathsf{\delta}_{\Omega2}^{Bell}=\mathsf{\delta}_{3}^{(0)}=0$, and
we obtain $\mathsf{\delta}_{D}^{(0)}=0$ for $D\geq2$ (from the nonexistent
$R^{D-2}$ term in Eq.(\ref{eq:QDser1})), whereas the first non null
curvature correction is $\mathsf{\delta}_{D}^{(1)}$ (from $R^{D-3}$
curvature term in Eq. (\ref{eq:QDser1})). To our best knowledge,
all this properties have been never studied or evaluated for $D\neq2$
and $3$ \cite{Bell,McQ,Soko,Urru}, and here we are showing the systematic
dimensional dependence in terms of the second cluster integral coefficient
in a higher dimensionality space (see Eq. (\ref{eq:delta1D})). Concisely,
Eq. (4.5) in III Ref. \cite{Bell} for a $D$ dimensional system must
be written as \begin{eqnarray}
\gamma/\gamma_{flat}-1 & = & -\mathrm{j}^{2}\frac{\mathsf{\delta}_{D}^{(1)}}{a_{D}}+O_{4}(R^{-1})+O(\rho)\, O(R^{-1})\:,\nonumber \\
 & \simeq & -\left(\frac{\sigma}{R}\right)^{2}\,\frac{(D-1)(D+1)}{2^{3}3\,(D+3)}\:.\label{eq:BellSurfT}\end{eqnarray}
where now $\gamma_{flat}$ is the surface tension of the open system
in contact with a planar wall. An interesting fact is that first correction
is then quadratic in curvature and zero order in density. Besides,
for a HS fluid in contact with a hard convex spherical wall (sometimes
referred in the literature as a spherical cavity inside of the bulk
fluid) this first order curvature correction should be exactly the
same. The term proportional to $-R^{-2}$ in Eq. (\ref{eq:BellSurfT})
is $1/40$ and $1/18$ for $D=2$ and $3$ respectively and becomes
greater than one for $D\geq27$. A consequence of Eq. (\ref{eq:BellSurfT})
is that the Tolman length of the athermal system of HS-HWSP scales
as $\delta^{Tol}\sim\rho$ for low density. It seems that the study
of the three particle system will provide the value of the proportionality
constant. We may conclude this section by noting that a curvature
correction proportional to $\sigma/R$ should exist at Eqs. (\ref{eq:SurfTser02},
\ref{eq:SurfTserN}) if we consider a somewhat more realistic potentials
with soft repulsion and/or attractive well.

Here, we have presented an study of the bulk properties of a system
consisting in 2HS-HWSP. In such an analysis we followed deliberately
a non-thermodynamic approach. Even when it may be unexpected, the
direct derivation of properties such as the pressure and surface tension
using a thermodynamic approach is not a simple task \cite{Urru2}.
The full implementation of this path requires a careful evaluation
of several problems related with the nonextensivity of the system.
This subject will be analyzed in an incoming work.

\section{Final remarks\label{sec:Conclusions}}

The few body system consisting of two hard spheres in a spherical
pore in any dimension has been studied in the framework of the canonical
ensemble of statistical mechanics. It was showed that several properties
of such an inhomogeneous spherical system can be exactly evaluated.
Analytical exact expressions for the canonical partition function
$\tilde{Q}_{D}$, density distribution function $\rho(r)$, and pressure
tensor $\mathbf{P}(r)$ were obtained. The iterative construction
of these functions in terms of the same properties for dimensions
lower than $D$ was performed. We should emphasize that neither approximations
nor power series truncation were done along these derivations. The
study of the analytical properties of CI at the low density limit
or large $R$ value, and the highly confinement limit or final solid
were analysed. We found that properties at both limits are correlated.
This becomes clear for odd $D$ values where the CI is a polynomial
and low density limit involves high order $R$ monomials, though caging
limit properties relates with the degree of the zero of CI at $R=\sigma/2$.
We found that such zero is of order $(D+3)/2$.

Other systems which are closely related to the 2-HS-HWSP have been
also analysed. The system of two sticky HS or rigid linear body into
a spherical pore was tackled. The two HS into a spherical pore with
a smaller internal and fixed hard core was studied with special emphasis
in the limit of on surface confinement. Several exact relations between
the three closely related systems were established by applying the
sticky bond transformation which makes possible their unified study.
It was examined the way in which properties of the three systems become
strongly dependent on the low density regime and on the opposite caging
limit. Several equations that relates the coefficients of these systems
were obtained.

The pressure tensor of the 2-HS-HWSP was investigated. The re-examination
of the mechanical equilibrium condition for a system with a hard spherical
boundary was done. New constraints between non ideal pressure tensor
components and one body distribution function slope at contact with
the curved wall was obtained. The analytical evaluation of one possible
definition of the pressure were performed. The obtained expression
disregards the former equilibrium condition and therefore the used
pressure tensor definition must be considered incorrect or at most
approximate.

The EOS of 2-HS-HWSP system was studied. We have analytically evaluated
the pressure-density relation and the surface tension. In connection
with the open system of HS in a HWSP, our results are consistent with
that obtained by Bellemans \cite{Bell} the first correction in the
surface tension due to the curvature in the confining surface is not
of order $R^{-1}$ in three dimension. We also show (see Eqs. (\ref{eq:QDser1},
\ref{eq:delta1D}) and (\ref{eq:BellSurfT})) that the first correction
has order $R^{-2}$ independently of the system's dimensionality and
is zero order in density. In addition we determined the value of the
coefficients corresponding to the first inhomogeneous and first non
planar wall corrections for all dimensions, Eqs. (\ref{eq:aD}, \ref{eq:delta1D}).
To the best of our knowledge, it is the first time that both coefficients
are evaluated.

The In-Out relation introduced in Ref. \cite{Urru} and described
in Section \ref{sec:Cap1}, the stick transformation introduced in
Section \ref{sec:Other-related-problems}, and several results (e.g.,
Eqs. (\ref{eq:QDser1}, \ref{eq:Qrigser1}, \ref{eq:Qg01}) and (\ref{eq:QDsser1}))
suggest interesting links with the mathematical theory of Convex Bodies
also known as Integral Geometry \cite{Meck,Muld} that will be studied
in future works.

\section*{Acknowledgments}

This work was supported in part by the Ministry of Culture and Education
of Argentina through Grants CONICET PIP No. 5138/05, ANPCyT PICT No
2006-00492 and UBACyT No. X298.

\appendix

\section{Appendix: Properties of $w_{D}(y,\Delta)$ \label{appendix-A}}

This appendix is devoted to the iterative construction of $w_{D}(y,\Delta)$.
The procedure to iteratively build $w_{D}(y,\Delta)$ is traced from
Eqs. (\ref{eq:wDrec}, \ref{eq:wyDdef}) where the last one applies
only for $\Delta\leq y\leq1$. We define the complementary function
of $w_{D}(y,\Delta)$ (see Eq. (\ref{eq:wyDdef})), $\widetilde{w}_{D}(y,\Delta)$
\begin{equation}
\widetilde{w}_{D}(y,\Delta)\equiv\frac{1}{2}\left(1+\Delta\right)^{D}w_{D}(x')-\frac{1}{2}\left(1-\Delta\right)^{D}w_{D}(x'')\:.\label{eqa:wDymdef}\end{equation}
with $x'=r'/2R_{1}=(y^{2}+\Delta)/[y\,(1+\Delta)]$ and $x''=r''/2R_{2}=(y^{2}-\Delta)/[y\,(1-\Delta)]$.
From the definition of $w_{D}(y,\Delta)$ (\ref{eq:wyDdef}), replacing
$w_{D}(x)$ with the aid of Eq. (\ref{eq:wDrec}) and rearranging
terms we obtain \begin{eqnarray}
w_{D}(y,\Delta) & = & (1+\Delta^{2})w_{D-2}(y,\Delta)+2\Delta\,\widetilde{w}_{D-2}(y,\Delta)-\nonumber \\
\nonumber \\ &  & \left(\left(\frac{R_{1}}{\bar{R}}\right)^{D}x'(1-x'^{2})^{(D-1)/2}+\left(\frac{R_{2}}{\bar{R}}\right)^{D}x''(1-x''^{2})^{(D-1)/2}\right)\mathtt{B}_{D}/D\:,\label{eqa:wyDrec00}\end{eqnarray}
Following a similar approach with Eq. (\ref{eqa:wDymdef}) and writing
$\{R_{1},R_{2}\}$ in terms of $y$ and $\Delta$ we obtain the iterative
relations \begin{eqnarray}
w_{D}(y,\Delta) & = & (1+\Delta^{2})w_{D-2}(y,\Delta)+2\Delta\,\widetilde{w}_{D-2}(y,\Delta)-\nonumber \\
\nonumber \\ &  & y^{2-D}\left((1-y^{2})(y^{2}-\Delta^{2})\right)^{(D-1)/2}2\mathtt{B}_{D}/D\:,\label{eqa:wyDrec}\end{eqnarray}
\begin{eqnarray}
\widetilde{w}_{D}(y,\Delta) & = & 2\Delta\, w_{D-2}(y,\Delta)+(1+\Delta^{2})\widetilde{w}_{D-2}(y,\Delta)-\nonumber \\
\nonumber \\ &  & \Delta\, y^{-D}\left((1-y^{2})(y^{2}-\Delta^{2})\right)^{(D-1)/2}2\mathtt{B}_{D}/D\:.\label{eqa:wymDrec}\end{eqnarray}
First functions of the series are\begin{equation}
w_{-1}(y,\Delta)=1/(1-\Delta^{2})\;,\label{eqa:wym1}\end{equation}
\begin{equation}
\widetilde{w}_{-1}(y,\Delta)=-\Delta/(1-\Delta^{2})\;,\label{eqa:wym1m}\end{equation}
\begin{equation}
w_{0}(y,\Delta)=\frac{1}{\pi}\,\left[arccos(\frac{y^{2}+\Delta}{y(1+\Delta)})+arccos(\frac{y^{2}-\Delta}{y(1-\Delta)})\right]\:,\label{eqa:wy0}\end{equation}
\begin{equation}
\widetilde{w}_{0}(y,\Delta)=\frac{1}{\pi}\,\left[arccos(\frac{y^{2}+\Delta}{y(1+\Delta)})-arccos(\frac{y^{2}-\Delta}{y(1-\Delta)})\right]\:,\label{eqa:wy0m}\end{equation}
\begin{equation}
w_{1}(y,\Delta)=1-y\:.\label{eqa:wDy}\end{equation}
\begin{equation}
\widetilde{w}_{1}(y,\Delta)=-\Delta\left(1-y\right)/y\:.\label{eqa:wDym}\end{equation}

\section{Appendix: Properties of $q_{D}(z)$ and $u_{D}(z)$\label{appendix-B}}

This appendix is devoted to deduct a few properties of functions $q_{D}(z)$
and $u_{D}(z)$. We begin summarizing four properties of $I_{x}(a,b)$
(from \cite{Functions} and \cite{BookAbram} p. 944)\begin{equation}
I_{x}(a,b)=1-I_{1-x}(b,a)\:,\label{eqa:Irec-id1}\end{equation}
\begin{equation}
\intop\, x^{c-1}I_{x}(a,b)\, dx=c^{-1}x^{c}I_{x}(a,b)-c^{-1}\frac{\Gamma(a+b)\Gamma(a+c)}{\Gamma(a)\Gamma(a+b+c)}I_{x}(a+c,b)\:,\label{eqa:Irec-id2}\end{equation}
\begin{equation}
I_{x}(a,b)=I_{x}(a+1,b)+(a\, B(a,b))^{-1}x^{a}(1-x)^{b}\:,\label{eqa:Irec-id3}\end{equation}
\begin{equation}
I_{x}(a,a)=2^{-1}I_{1-4(x-1/2)^{2}}(a,1/2)\;,\, with\, x\leq1/2\:.\label{eqa:Irec-id4}\end{equation}
The deduction of Eqs. (\ref{eq:uDef}, \ref{eq:uiter}) for $q(z)$
follows from the definition on Eq. (\ref{eq:QD_2})

\begin{equation}
q_{D}(z)\equiv2^{D}D\,\int_{z}^{1}\, x^{D-1}\, w_{D}(x)\, dx\:,\label{eqa:qDdef}\end{equation}
where $0<z<1$, which may be rearranged using Eqs. (\ref{eq:wDI},
\ref{eqa:Irec-id1}) \begin{eqnarray}
q_{D}(z) & = & 2^{D}D\,\int_{z}^{1}\, t^{D-1}\, I_{1-t^{2}}((D+1)/2,1/2)\, dt\:,\nonumber \\
 & = & 2^{D}\,(1-z^{D})-2^{D-1}D\,\int_{z^{2}}^{1}\, t^{D/2-1}I_{t}(1/2,(D+1)/2)\, dt\:,\label{eqa:qD001}\end{eqnarray}
integrating on through identity (\ref{eqa:Irec-id2}), and using Eq.
(\ref{eqa:Irec-id1}) and definitions of Eqs. (\ref{eq:wDI}, \ref{eq:uDef})
\begin{eqnarray}
q_{D}(z) & \begin{array}{c}
=\end{array} & 2^{D}(1-z^{D})-2^{D}\left.\left(t^{D/2}I_{t}(1/2,(D+1)/2)-2^{-D}I_{t}((D+1)/2),(D+1)/2)\right)\right|_{z^{2}}^{1}\nonumber \\
 & = & 2^{D}(1-z^{D})-2^{D}\, I_{1}(1/2,(D+1)/2)+(2z)^{D}I_{z^{2}}(1/2,(D+1)/2)+\nonumber \\
 &  & I_{1}((D+1)/2,(D+1)/2)-I_{z^{2}}((D+1)/2,(D+1)/2)\nonumber \\
 & = & -(2z)^{D}I_{1-z^{2}}((D+1)/2,1/2)+I_{1-z^{2}}((D+1)/2,(D+1)/2)\nonumber \\
 & = & u_{D}(z)-(2z)^{D}w_{D}(z)\:.\label{eqa:qD002}\end{eqnarray}
Interestingly, above expressions may transform to show a complete
dependence on $w_{D}(z)$. Using the identity (\ref{eqa:Irec-id4})
and the analytic extension of $w_{D}(z)$ (\ref{eq:ZD}) we find $u_{D}(z)=2^{-1}w_{D}(z')$
with $z'=2z^{2}-1$, and\begin{equation}
q_{D}(z)=2^{-1}w_{D}(z')-(2z)^{D}w_{D}(z)\:,\label{eq:qDwD}\end{equation}
therefore, $q_{D}(z)$ may be built in terms of the recurrence relation
Eq. (\ref{eq:wDrec}). With the purpose of deduce the Eq. (\ref{eq:uiter})
we will find some identities not available in the literature of Refs.
\cite{BookAbram,Functions}. From (\ref{eqa:Irec-id1}, \ref{eqa:Irec-id3})
we obtain\begin{equation}
I_{1-x}(a,b)=I_{1-x}(a,b+1)-(b\, B(a,b))^{-1}x^{b}(1-x)^{a}\:,\label{eqa:Irec4}\end{equation}
\begin{equation}
I_{1-x}(a,b)=I_{1-x}(a+1,b)+(a\, B(a,b))^{-1}x^{b}(1-x)^{a}\:,\label{eq:Irec5}\end{equation}
applying both recurrence relations we find\begin{equation}
I_{1-x}(a+1,b+1)=I_{1-x}(a,b)+(a\, b\, B(a,b)/(a+b))^{-1}x^{b}(1-x)^{a}(\frac{a}{a+b}-x)\:,\label{eqa:Irec6}\end{equation}
\begin{equation}
I_{1-x}(a+1,a+1)=I_{1-x}(a,a)+\frac{2\Gamma(2a)}{a\Gamma^{2}(a)}x^{a}(1-x)^{a}(\frac{1}{2}-x)\:,\label{eqa:Irec7}\end{equation}
and then using Eq. (\ref{eq:uDef})\begin{equation}
u_{D}(z)=u_{D-2}(z)+\frac{\Gamma(D)}{\Gamma^{2}((D+1)/2)}z^{D-1}(1-z^{2})^{(D-1)/2}(\frac{1}{2}-z^{2})\:.\label{eqa:uDrec}\end{equation}
 From Eq. (\ref{eq:uDef}) the first functions $u_{D}(z)$ of the
series may be obtained (see Eqs. (\ref{eq:um1}, \ref{eq:u0})). Finally,
we may mention that $q_{D}(z)$ could be also defined as the solution
of the second order differential equation\begin{equation}
\partial_{z}^{2}q_{D}-(D-1)\, z^{-1}\partial_{z}q_{D}=2^{D-1}\mathtt{B}_{D}D\,\left(z^{2}(1-z^{2})\right)^{(D-1)/2}\:,\label{eq:qDdifeq}\end{equation}
with boundary conditions $q_{D}(0)=1$ and $q_{D}(1)=0$.

\end{document}